\pdfoutput=1
\documentclass[11pt,a4paper]{article}

\usepackage{ifthen} 
\newboolean{pdflatex}
\setboolean{pdflatex}{true} % False for eps figures 
\newboolean{articletitles}
\setboolean{articletitles}{true} % False removes titles in references
\newboolean{uprightparticles}
\setboolean{uprightparticles}{false} %True for upright particle symbols
\newboolean{inbibliography}
\setboolean{inbibliography}{false} %True once you enter the bibliography

\usepackage{url}

\newcommand{\IF}[4]{\ifthenelse{\equal{#1}{#2}}{#3}{#4}}%
\usepackage[top=1in, bottom=1.25in, left=1in, right=1in]{geometry}

\usepackage{microtype}
\usepackage{lineno}  % for line numbering during review
\usepackage{xspace} % To avoid problems with missing or double spaces after
\usepackage{caption} %these three command get the figure and table captions automatically small

\usepackage{graphicx}  % to include figures (can also use other packages)
\usepackage{color}
\usepackage{colortbl}
\graphicspath{{./figs/}} % Make Latex search fig subdir for figures

\usepackage{amsmath} % Adds a large collection of math symbols
\usepackage{amssymb}
\usepackage{amsfonts}
\usepackage{upgreek} % Adds in support for greek letters in roman typeset

\newcommand*\patchAmsMathEnvironmentForLineno[1]{%
\expandafter\let\csname old#1\expandafter\endcsname\csname #1\endcsname
\expandafter\let\csname oldend#1\expandafter\endcsname\csname
end#1\endcsname
 \renewenvironment{#1}%
   {\linenomath\csname old#1\endcsname}%
   {\csname oldend#1\endcsname\endlinenomath}%
}
\newcommand*\patchBothAmsMathEnvironmentsForLineno[1]{%
  \patchAmsMathEnvironmentForLineno{#1}%
  \patchAmsMathEnvironmentForLineno{#1*}%
}
\AtBeginDocument{%
\patchBothAmsMathEnvironmentsForLineno{equation}%
\patchBothAmsMathEnvironmentsForLineno{align}%
\patchBothAmsMathEnvironmentsForLineno{flalign}%
\patchBothAmsMathEnvironmentsForLineno{alignat}%
\patchBothAmsMathEnvironmentsForLineno{gather}%
\patchBothAmsMathEnvironmentsForLineno{multline}%
\patchBothAmsMathEnvironmentsForLineno{eqnarray}%
}

\usepackage{mciteplus}

\def\aerr#1#2{{{#1}\atop{#2}}}

\usepackage{xspace} 
\usepackage{upgreek}

\def\dzero  {\mbox{D0}\xspace}

\def\MagUp {\mbox{\em Mag\kern -0.05em Up}\xspace}

\ifthenelse{\boolean{uprightparticles}}%
{

 \def\Pmu         {\ensuremath{\upmu}\xspace}                 
 \def\Pnu         {\ensuremath{\upnu}\xspace}                 
                  
 \def\Ppi         {\ensuremath{\uppi}\xspace}                 
                  
 \def\Prho        {\ensuremath{\uprho}\xspace}                 
                  
 \def\Ptau        {\ensuremath{\uptau}\xspace}

 \def\Ppsi        {\ensuremath{\uppsi}\xspace}

 \def\PDelta      {\ensuremath{\Delta}\xspace}                 
 \def\PXi      {\ensuremath{\Xi}\xspace}                 
 \def\PLambda      {\ensuremath{\Lambda}\xspace}                 
 \def\PSigma      {\ensuremath{\Sigma}\xspace}                 
 \def\POmega      {\ensuremath{\Omega}\xspace}                 
 \def\PUpsilon      {\ensuremath{\Upsilon}\xspace}                 
 
 \def\PB      {\ensuremath{\mathrm{B}}\xspace}                 
                  
 \def\PD      {\ensuremath{\mathrm{D}}\xspace}

 \def\PJ      {\ensuremath{\mathrm{J}}\xspace}                 
 \def\PK      {\ensuremath{\mathrm{K}}\xspace}

 \def\PW      {\ensuremath{\mathrm{W}}\xspace}

 \def\Pb      {\ensuremath{\mathrm{b}}\xspace}                 
 \def\Pc      {\ensuremath{\mathrm{c}}\xspace}                 
 \def\Pd      {\ensuremath{\mathrm{d}}\xspace}                 
 \def\Pe      {\ensuremath{\mathrm{e}}\xspace}

 \def\Pi      {\ensuremath{\mathrm{i}}\xspace}

 \def\Pp      {\ensuremath{\mathrm{p}}\xspace}                 
 \def\Pq      {\ensuremath{\mathrm{q}}\xspace}                 
                  
 \def\Ps      {\ensuremath{\mathrm{s}}\xspace}                 
 \def\Pt      {\ensuremath{\mathrm{t}}\xspace}                 
 \def\Pu      {\ensuremath{\mathrm{u}}\xspace}

}
{

 \def\Pmu         {\ensuremath{\mu}\xspace}                 
 \def\Pnu         {\ensuremath{\nu}\xspace}                 
                  
 \def\Ppi         {\ensuremath{\pi}\xspace}                 
                  
 \def\Prho        {\ensuremath{\rho}\xspace}                 
                  
 \def\Ptau        {\ensuremath{\tau}\xspace}

 \def\Ppsi        {\ensuremath{\psi}\xspace}                 
                  
 \mathchardef\PDelta="7101
 \mathchardef\PXi="7104
 \mathchardef\PLambda="7103
 \mathchardef\PSigma="7106
 \mathchardef\POmega="710A
 \mathchardef\PUpsilon="7107
                  
 \def\PB      {\ensuremath{B}\xspace}                 
                  
 \def\PD      {\ensuremath{D}\xspace}

 \def\PJ      {\ensuremath{J}\xspace}                 
 \def\PK      {\ensuremath{K}\xspace}

 \def\PW      {\ensuremath{W}\xspace}

 \def\Pb      {\ensuremath{b}\xspace}                 
 \def\Pc      {\ensuremath{c}\xspace}                 
 \def\Pd      {\ensuremath{d}\xspace}                 
 \def\Pe      {\ensuremath{e}\xspace}

 \def\Pi      {\ensuremath{i}\xspace}

 \def\Pp      {\ensuremath{p}\xspace}                 
 \def\Pq      {\ensuremath{q}\xspace}                 
                  
 \def\Ps      {\ensuremath{s}\xspace}                 
 \def\Pt      {\ensuremath{t}\xspace}                 
 \def\Pu      {\ensuremath{u}\xspace}

}

\makeatletter
\ifcase \@ptsize \relax% 10pt
  \newcommand{\miniscule}{\@setfontsize\miniscule{4}{5}}% \tiny: 5/6
\or% 11pt
  \newcommand{\miniscule}{\@setfontsize\miniscule{5}{6}}% \tiny: 6/7
\or% 12pt
  \newcommand{\miniscule}{\@setfontsize\miniscule{5}{6}}% \tiny: 6/7
\fi
\makeatother

\DeclareRobustCommand{\optbar}[1]{\shortstack{{\miniscule (\rule[.5ex]{1.25em}{.18mm})}
  \\ [-.7ex] $#1$}}

\def\epem       {{\ensuremath{\Pe^+\Pe^-}}\xspace}

\def\mun        {{\ensuremath{\Pmu^-}}\xspace} % muon negative (\mum is taken)
\def\mumu       {{\ensuremath{\Pmu^+\Pmu^-}}\xspace}

\def\ellm       {{\ensuremath{\ell^-}}\xspace}

\def\ellell     {\ensuremath{\ell^+ \ell^-}\xspace}

\def\neub       {{\ensuremath{\overline{\Pnu}}}\xspace}
\def\W      {{\ensuremath{\PW}}\xspace}

\def\quark     {{\ensuremath{\Pq}}\xspace}
\def\quarkbar  {{\ensuremath{\overline \quark}}\xspace}

\def\uquark    {{\ensuremath{\Pu}}\xspace}

\def\dquark    {{\ensuremath{\Pd}}\xspace}

\def\squark    {{\ensuremath{\Ps}}\xspace}
\def\squarkbar {{\ensuremath{\overline \squark}}\xspace}

\def\cquark    {{\ensuremath{\Pc}}\xspace}
\def\cquarkbar {{\ensuremath{\overline \cquark}}\xspace}

\def\bquark    {{\ensuremath{\Pb}}\xspace}
\def\bquarkbar {{\ensuremath{\overline \bquark}}\xspace}

\def\tquark    {{\ensuremath{\Pt}}\xspace}

\def\pion   {{\ensuremath{\Ppi}}\xspace}
\def\piz    {{\ensuremath{\pion^0}}\xspace}

\def\pip    {{\ensuremath{\pion^+}}\xspace}
\def\pim    {{\ensuremath{\pion^-}}\xspace}
\def\pipm   {{\ensuremath{\pion^\pm}}\xspace}

\def\rhomeson {{\ensuremath{\Prho}}\xspace}
\def\rhoz     {{\ensuremath{\rhomeson^0}}\xspace}

\def\kaon    {{\ensuremath{\PK}}\xspace}
  \def\Kbar    {{\kern 0.2em\overline{\kern -0.2em \PK}{}}\xspace}

\def\KorKbar    {\kern 0.18em\optbar{\kern -0.18em K}{}\xspace}
\def\Kz      {{\ensuremath{\kaon^0}}\xspace}

\def\Kp      {{\ensuremath{\kaon^+}}\xspace}
\def\Km      {{\ensuremath{\kaon^-}}\xspace}

\def\Kmp     {{\ensuremath{\kaon^\mp}}\xspace}
\def\KS      {{\ensuremath{\kaon^0_{\mathrm{ \scriptscriptstyle S}}}}\xspace}
\def\KL      {{\ensuremath{\kaon^0_{\mathrm{ \scriptscriptstyle L}}}}\xspace}
\def\Kstarz  {{\ensuremath{\kaon^{*0}}}\xspace}
\def\Kstarzb {{\ensuremath{\Kbar{}^{*0}}}\xspace}

  \def\Dbar    {{\kern 0.2em\overline{\kern -0.2em \PD}{}}\xspace}
\def\D       {{\ensuremath{\PD}}\xspace}

\def\DorDbar    {\kern 0.18em\optbar{\kern -0.18em D}{}\xspace}
\def\Dz      {{\ensuremath{\D^0}}\xspace}

\def\Dp      {{\ensuremath{\D^+}}\xspace}

\def\Ds      {{\ensuremath{\D^+_\squark}}\xspace}

\def\Dspm    {{\ensuremath{\D^{\pm}_\squark}}\xspace}

\def\B       {{\ensuremath{\PB}}\xspace}
\def\Bbar    {{\ensuremath{\kern 0.18em\overline{\kern -0.18em \PB}{}}}\xspace}

\def\BorBbar    {\kern 0.18em\optbar{\kern -0.18em B}{}\xspace}
\def\Bz      {{\ensuremath{\B^0}}\xspace}
\def\Bzb     {{\ensuremath{\Bbar{}^0}}\xspace}
\def\Bu      {{\ensuremath{\B^+}}\xspace}

\def\Bp      {{\ensuremath{\Bu}}\xspace}

\def\Bd      {{\ensuremath{\B^0}}\xspace}
\def\Bs      {{\ensuremath{\B^0_\squark}}\xspace}
\def\Bsb     {{\ensuremath{\Bbar{}^0_\squark}}\xspace}
\def\Bdb     {{\ensuremath{\Bbar{}^0}}\xspace}

\def\jpsi     {{\ensuremath{{\PJ\mskip -3mu/\mskip -2mu\Ppsi\mskip 2mu}}}\xspace}

  \def\Y#1S{\ensuremath{\PUpsilon{(#1S)}}\xspace}% no space before {...}!

\def\proton      {{\ensuremath{\Pp}}\xspace}

\def\Lz          {{\ensuremath{\PLambda}}\xspace}
\def\Lbar        {{\ensuremath{\kern 0.1em\overline{\kern -0.1em\PLambda}}}\xspace}
\def\LorLbar    {\kern 0.18em\optbar{\kern -0.18em \PLambda}{}\xspace}

\def\Lb      {{\ensuremath{\Lz^0_\bquark}}\xspace}

\def\Lc      {{\ensuremath{\Lz^+_\cquark}}\xspace}

\def\BF         {{\ensuremath{\mathcal{B}}}\xspace}

\def\BR         {\BF}
\newcommand{\decay}[2]{\ensuremath{#1\!\to #2}\xspace}         % {\Pa}{\Pb \Pc}

\def\to                 {\ensuremath{\rightarrow}\xspace}

\def\CP                {{\ensuremath{C\!P}}\xspace}

\def\Vud  {{\ensuremath{V_{\uquark\dquark}}}\xspace}
\def\Vcd  {{\ensuremath{V_{\cquark\dquark}}}\xspace}
\def\Vtd  {{\ensuremath{V_{\tquark\dquark}}}\xspace}
\def\Vus  {{\ensuremath{V_{\uquark\squark}}}\xspace}
\def\Vcs  {{\ensuremath{V_{\cquark\squark}}}\xspace}
\def\Vts  {{\ensuremath{V_{\tquark\squark}}}\xspace}
\def\Vub  {{\ensuremath{V_{\uquark\bquark}}}\xspace}
\def\Vcb  {{\ensuremath{V_{\cquark\bquark}}}\xspace}
\def\Vtb  {{\ensuremath{V_{\tquark\bquark}}}\xspace}

\def\Vcbs  {{\ensuremath{V_{\cquark\bquark}^\ast}}\xspace}
\def\Vtbs  {{\ensuremath{V_{\tquark\bquark}^\ast}}\xspace}

\newcommand{\dm}{{\ensuremath{\Delta m}}\xspace}

\def\AT#1     {\ensuremath{A_{\mathrm{T}}^{#1}}\xspace}           % 2

\def\C#1      {\ensuremath{\mathcal{C}_{#1}}\xspace}                       % 9
\def\Cp#1     {\ensuremath{\mathcal{C}_{#1}^{'}}\xspace}                    % 7
\def\Ceff#1   {\ensuremath{\mathcal{C}_{#1}^{\mathrm{(eff)}}}\xspace}        % 9  
\def\Cpeff#1  {\ensuremath{\mathcal{C}_{#1}^{'\mathrm{(eff)}}}\xspace}       % 7
\def\Ope#1    {\ensuremath{\mathcal{O}_{#1}}\xspace}                       % 2
\def\Opep#1   {\ensuremath{\mathcal{O}_{#1}^{'}}\xspace}                    % 7

\newcommand{\tev}{\ifthenelse{\boolean{inbibliography}}{\ensuremath{~T\kern -0.05em eV}}{\ensuremath{\mathrm{\,Te\kern -0.1em V}}}\xspace}
\newcommand{\gev}{\ensuremath{\mathrm{\,Ge\kern -0.1em V}}\xspace}
\newcommand{\mev}{\ensuremath{\mathrm{\,Me\kern -0.1em V}}\xspace}
\newcommand{\kev}{\ensuremath{\mathrm{\,ke\kern -0.1em V}}\xspace}
\newcommand{\ev}{\ensuremath{\mathrm{\,e\kern -0.1em V}}\xspace}
\newcommand{\gevc}{\ensuremath{{\mathrm{\,Ge\kern -0.1em V\!/}c}}\xspace}
\newcommand{\mevc}{\ensuremath{{\mathrm{\,Me\kern -0.1em V\!/}c}}\xspace}
\newcommand{\gevcc}{\ensuremath{{\mathrm{\,Ge\kern -0.1em V\!/}c^2}}\xspace}
\newcommand{\gevgevcccc}{\ensuremath{{\mathrm{\,Ge\kern -0.1em V^2\!/}c^4}}\xspace}
\newcommand{\mevcc}{\ensuremath{{\mathrm{\,Me\kern -0.1em V\!/}c^2}}\xspace}

\def\invps{\ensuremath{{\mathrm{ \,ps^{-1}}}}\xspace}
\def\invns{\ensuremath{{\mathrm{ \,ns^{-1}}}}\xspace}

\def\gsim{{~\raise.15em\hbox{$>$}\kern-.85em
          \lower.35em\hbox{$\sim$}~}\xspace}
\def\lsim{{~\raise.15em\hbox{$<$}\kern-.85em
          \lower.35em\hbox{$\sim$}~}\xspace}

\def\degrees{\ensuremath{^{\circ}}\xspace}

\def\tell1  {TELL1\xspace}
\def\ukl1   {UKL1\xspace}

\usepackage{rotating}

\def\dzero  {\mbox{D\O}\xspace}

\newcommand{\boldsubsection}[1]{\subsection[#1]{\boldmath #1}}
\newcommand{\boldsubsubsection}[1]{\subsubsection[#1]{\boldmath #1}}

{
\topmargin = -12mm
}
\usepackage{hyperref}
\usepackage[all]{hypcap} % Internal hyperlinks to floats.
\usepackage{hyperxmp}
\usepackage{cite} % Citation ranges. Last thing before begin document.
\begin{document}

\begin{titlepage}

% Header ---------------------------------------------------
\vspace*{-1.5cm}

\noindent
\begin{tabular*}{\linewidth}{lc@{\extracolsep{\fill}}r@{\extracolsep{0pt}}}
\ifthenelse{\boolean{pdflatex}}% Logo format choice
 & & Nikhef-2017-012 \\  % ID 
 & & LPT-Orsay-17-06  \\
 & & \today \\ % Date - Can also hardwire e.g.: 23 March 2010
\hline
\end{tabular*}

\vspace*{4.0cm}

% Title --------------------------------------------------
{\bf\boldmath\huge
\begin{center}
The CKM Parameters
\end{center}
}

\vspace*{2.0cm}

% Authors -------------------------------------------------
\begin{center}
S\'ebastien Descotes-Genon$^1$ and Patrick Koppenburg$^2$
\bigskip\\
{\normalfont\itshape\footnotesize
$^1$Laboratoire de Physique Th\'eorique  (UMR 8627),
  CNRS, Univ. Paris-Sud,\\ Universit\'e Paris-Saclay;\\ email: sebastien.descotes-genon@th.u-psud.fr\\
  \vskip 0.5em
$^2$Nikhef, 1098 XG Amsterdam;\\ email: patrick.koppenburg@nikhef.nl
}
\end{center}

\vspace{\fill}

% Abstract -----------------------------------------------
\begin{abstract}
  The Cabibbo--Kobayashi--Maskawa (CKM)\ matrix is a key element in describing
flavour dynamics in the Standard Model. With only four parameters,
this matrix is able to describe a large range of phenomena in the
quark sector, such as \CP violation and rare decays. It can thus
be constrained by many different processes, which have to be
measured experimentally with high accuracy and computed with good theoretical control. Recently, with the advent of the \B factories and
the LHCb experiment taking data, the precision has significantly
improved. We review the most relevant experimental constraints and
theoretical inputs and present fits to the CKM matrix for the Standard Model and for some topical
model-independent studies of New Physics.

\end{abstract}

\vspace*{2.0cm}
\centerline{Invited contribution to Annual Review of Nuclear and Particle Science, 67:97-127.}
\vspace{\fill}
\vspace*{2mm}
\end{titlepage}

\pagestyle{empty}  % no page number for the title 

%%%%%%%%%%%%%%%%%%%%%%%%%%%%%%%%
%%%%%  EOD OF TITLE PAGE  %%%%%%
%%%%%%%%%%%%%%%%%%%%%%%%%%%%%%%%

\tableofcontents
\cleardoublepage
\pagestyle{plain}
\pagenumbering{arabic}

\section{INTRODUCTION}\label{Sec:Introduction}

The study of elementary particles and their electromagnetic, weak
and strong interactions has led to a particularly successful
theory, the Standard Model (SM). The SM has been extensively tested,
culminating with the recent discovery of the Higgs
boson~\cite{Aad:2012tfa,Chatrchyan:2012ufa} at the Large Hadron
Collider (LHC). In the development of this description, quark
flavour physics has played a central role in two different
aspects. First, the SM embeds the Kobayashi--Maskawa mechanism: The
Cabibbo--Kobayashi--Maskawa (CKM) mixing
matrix~\cite{Cabibbo:1963yz,Kobayashi:1973fv} arising in charged
weak interactions represents the single source of all observed
differences between particles and antiparticles, namely \CP
violation in the quark sector. Second, flavour-changing currents
(in particular, neutral ones) have repeatedly revealed evidence for
new, heavier degrees of freedom (charm quark, weak gauge bosons,
top quark) before their discovery.

Yet the SM fails in some key aspects. Why is there such a large
number of parameters for quark masses and the CKM mixing matrix,
spanning such a wide range of values? Why are the electroweak and
strong interactions treated separately? Why is antimatter absent from the observed universe, even though the amount of \CP
violation in the SM is too small to produce the observed
matter--antimatter
asymmetry~\cite{Sakharov:1967dj,Cohen:1993nk,Riotto:1999yt,Hou:2008xd}? New Physics (NP) extensions of the SM are expected to address
these issues by including heavier particles related to
higher-energy phenomena. The related shorter-distance interactions
would have immediate consequences not only in production
experiments at high energies but also through deviations from the
SM predictions in flavour processes (new sources of \CP violation,
interferences between SM and NP contributions).

Therefore, a precision study of the CKM matrix is certainly
desirable from a practitioner's point of view: Performing the
metrology of the SM parameters yields accurate predictions for
weak transitions, including \CP-violating processes. But it is
also required from a more theoretical point of view: The mixing
due to the CKM matrix in weak processes has a very simple and
constrained structure in the SM and is generally affected
significantly by NP extensions, constituting a very powerful probe
of models beyond the SM. The need for an accurate determination of the CKM matrix has led to an impressive effort from
the experimental community, specifically the extensive research performed at
the BaBar and Belle experiments, the large data samples available
at the LHC, and the advent of the high-luminosity Belle-II \B
factory. The theoretical community has also made remarkable
progress in the understanding of strong and weak interactions of
the quarks, both analytically (in particular, through the
development of effective theories) and numerically (with improvements in lattice simulations of QCD). Thus, very high
precision measurements of CKM parameters are both needed and
currently accessible, and they are the object of this review. We discuss the
theoretical grounds related to the CKM matrix in Section 2, review the main experimental constraints on its
parameters in Section~\ref{Sec:Constraints}, and
present examples of global analyses of the CKM matrix and the impact of NP
contributions in Section~\ref{Sec:global}.

\section{THE CKM MATRIX}\label{Sec:CKM}

\boldsubsection{Structure of the CKM Matrix}\label{Sec:CKMstruct}

In the SM, the Lagrangian for the Yukawa coupling of the Higgs boson to the quark
fields yields (after electroweak symmetry breaking)
\begin{equation}
\mathcal{L}^{q}_M=-(M_d)_{ij} \overline{D'_{{L}i}} D'_{{R}j}
-(M_u)_{ij} \overline{U'_{{L}i}} U'_{{R}j},
\end{equation}
where $i$ and $j$ are family indices, with $U'=(u',\,c',\,t')$ and 
$D=(d',\,s',\,b')$, and L and R indicate the components with left-
and right-handed chiralities, respectively. The prime symbols indicate that these fields
are not necessarily the mass eigenstates of the theory. The
matrices $M_u$ and $M_d$ are related to the Yukawa coupling
matrices as $M_q=vY^q/\sqrt{2}$, where $v$ is the vacuum expectation
value (the neutral component) of the Higgs field. At this
stage, $M_u$ and $M_d$ are general complex matrices to be
diagonalised using the singular value decomposition $M_q=V^\dag_{q{L}}
m_q V_{q{R}}$, where $V_{{L},{R}}$ is  unitary and $m_q$ is diagonal, real, and
positive. The mass eigenstates are identified as $U_{L}=V_{u{L}}
U'_{L}$ and $U_{R}=V_{u{R}} U'_{R}$, and similarly for $D$.

Expressing the interactions of quarks with gauge bosons in terms
of mass eigenstates does not modify the structure of the
Lagrangian in the case of neutral gauge bosons, but it affects
charged-current interactions between quarks and $W^\pm$, described by the Lagrangian
\begin{equation}\label{eq:chargeInteraction}
\mathcal{L}_{W^{\pm}} = -
\frac{g}{\sqrt{2}}\overline{U}_{i}\gamma^{\mu}\frac{1-\gamma^5}{2}\left(V_{\rm
CKM} \right )_{ij} D_{j} W_{\mu}^{+} + \mathrm{h.c.},
\end{equation}
where $g$ is the electroweak coupling constant and $V_{\rm
CKM}=V_{u{L}}^\dag V_{d{L}}$ is the unitary CKM matrix:

\begin{equation}\label{eq:ckmMatrix}
V_{\rm CKM} = \left(
\begin{array}{lcr}
V_{ud} & V_{us} & V_{ub} \\
V_{cd} & V_{cs} & V_{cb} \\
V_{td} & V_{ts} & V_{tb}
\end{array}
\right).
\end{equation}
The CKM matrix induces flavour-changing transitions inside and between generations in the charged sector at tree level ($W^\pm$ interaction). By contrast, there are no flavour-changing transitions in the neutral sector at tree level ($Z^0$ and photon interactions). The CKM matrix stems from the Yukawa interaction between the Higgs boson and the fermions, and it originates from the misalignment in flavour space of the up and
down components of the $SU(2)_\mathrm{L}$ quark doublets of the SM (as
there is no dynamical mechanism in the SM to enforce
$V_{u{L}}=V_{d{L}}$). The $V_{{\rm CKM},ij}$ CKM matrix elements
(hereafter, $V_{ij}$) represent the couplings between up-type
quarks $U_i=(u,\,c,\,t)$ and down-type quarks $D_j=(d,\,s,\,b)$.
There is some arbitrariness in the conventions used to define this matrix.
In particular, the relative phases among the left-handed quark
fields can be redefined, reducing the number of real parameters
describing this unitary matrix from three moduli and six phases to
three moduli and one phase [more generally, for $N$ generations, one has $N(N-1)/2$ moduli and $(N-1)(N-2)/2$ phases]. Because \CP conjugate processes correspond to
interaction terms in the Lagrangian related by Hermitian
conjugation, the presence of a phase, and thus the complex nature
of the CKM matrix, may induce differences between rates of \CP
conjugate processes, leading to \textit{CP} violation. This does not occur for only two generations, where $V_\mathrm{CKM}$ is real and parametrised by a single real parameter, the Cabibbo angle.

According to experimental evidence, transitions within the same
generation are characterised by $V_{\rm CKM}$ elements of
$\mathcal{O}(1)$. Those between the first and second generations
are suppressed by a factor of $\mathcal{O}(10^{-1})$; those between
the second and third generations by a factor of
$\mathcal{O}(10^{-2})$; and those between the first and third
generations by a factor of $\mathcal{O}(10^{-3})$. This hierarchy can
be expressed by defining the four phase convention--independent
quantities as follows:

\begin{equation}
\lambda^2§=\frac{|V_{us}|^2}{|V_{ud}|^2+|V_{us}|^2}\,, \qquad
A^2\lambda^4=\frac{|V_{cb}|^2}{|V_{ud}|^2+|V_{us}|^2}\,, \qquad
\bar\rho+i\bar\eta=-\frac{V_{ud}V_{ub}^*}{V_{cd}V^*_{cb}}\,.
\end{equation}
An alternative convention exists in the literature for the last
two CKM parameters, corresponding to

\begin{equation}
\rho+i\eta=\frac{V_{ub}^*}{V_{us}V^*_{cb}}
=\left(1+\frac{1}{2}\lambda^2\right)(\bar\rho+i\bar\eta)+O(\lambda^4).
\end{equation}

The CKM matrix can be expanded in powers of the small parameter
$\lambda$ (which corresponds to $\sin{\theta_{C}}\simeq
0.22$)~\cite{Wolfenstein:1983yz}, exploiting the unitarity of
$V_\text{CKM}$ to highlight its hierarchical structure. This
expansion yields the following parametrisation of the CKM\ matrix up to
$\mathcal{O}\left(\lambda^{6}\right)$:

\begin{equation}\label{eq:ckmWolfenstein4}
V_{\rm CKM} = \left(
\begin{array}{ccc}
1-\frac{1}{2}\lambda^{2}-\frac{1}{8}\lambda^{4} & \lambda & A\lambda^{3}\left(\bar\rho -i\bar\eta\right) \\
-\lambda +\frac{1}{2}A^{2}\lambda^{5}\left[1-2(\bar\rho +i\bar\eta)\right] & 1-\frac{1}{2}\lambda^{2}-\frac{1}{8}\lambda^{4}(1+4A^{2}) & A\lambda^{2} \\
A\lambda^{3}\left[1-(\bar\rho +i\bar\eta)\right]~ &
~-A\lambda^{2}+\frac{1}{2}A\lambda^{4}\left[1-2(\bar\rho
+i\bar\eta)\right]~ & ~1-\frac{1}{2}A^{2}\lambda^{4}
\end{array}
\right).
\end{equation}
The CKM matrix is complex; thus, \CP violation is allowed if
and only if $\bar\eta$ differs from zero. To lowest order, the
Jarlskog parameter measuring \CP violation in a
convention-independent manner~\cite{Jarlskog:1985ht},

\begin{equation}\label{eq:jarlskogValueWolf}
J_{C\!P} \equiv \left|\Im\left(V_{i\alpha} V_{j\beta}
V_{i\beta}^{*}V_{j\alpha}^{*}\right)\right| =
\lambda^{6}A^{2}\bar\eta, \qquad \left( i\neq j,
\alpha\neq\beta\right),
\end{equation}
is directly related to the \CP-violating parameter $\bar\eta$, as
expected.

\boldsubsection{The Unitarity Triangle}\label{Sec:UT}

To represent the knowledge of the four CKM parameters, it is
useful to exploit the unitarity condition of the CKM matrix:
$V_{\rm CKM}V_{\rm CKM}^{\dagger} = V_{\rm CKM}^{\dagger}V_{\rm
CKM} = \mathbb{I}$. This condition corresponds to a set of 12 equations: six
for diagonal terms and six for off-diagonal terms. In particular,
the equations for the off-diagonal terms can be represented as
triangles in the complex plane, all characterised by the same area
$J_{C\!P}/2$. Only two of these six triangles have sides of
the same order of magnitude, $\mathcal{O}(\lambda^{3})$ (i.e., are
not squashed):

\begin{equation}\label{eq:nonsquashedUT}
\underbrace{V_{ud}V_{ub}^{*}}_{\mathcal{O}(\lambda^{3})}+\underbrace{V_{cd}V_{cb}^{*}}_{\mathcal{O}(\lambda^{3})}+\underbrace{V_{td}V_{tb}^{*}}_{\mathcal{O}(\lambda^{3})}
= 0, \qquad
\underbrace{V_{ud}V_{td}^{*}}_{\mathcal{O}(\lambda^{3})}+\underbrace{V_{us}V_{ts}^{*}}_{\mathcal{O}(\lambda^{3})}+\underbrace{V_{ub}V_{tb}^{*}}_{\mathcal{O}(\lambda^{3})}
= 0.
\end{equation}
%%%%%%%%%%%%%%%%%%%%%%%%%%%%%%%%%%%%%%%%%%%%%%%%%%%%%%%%%%%%%%%%%%%
\begin{figure}\centering
\includegraphics[width=1\textwidth]{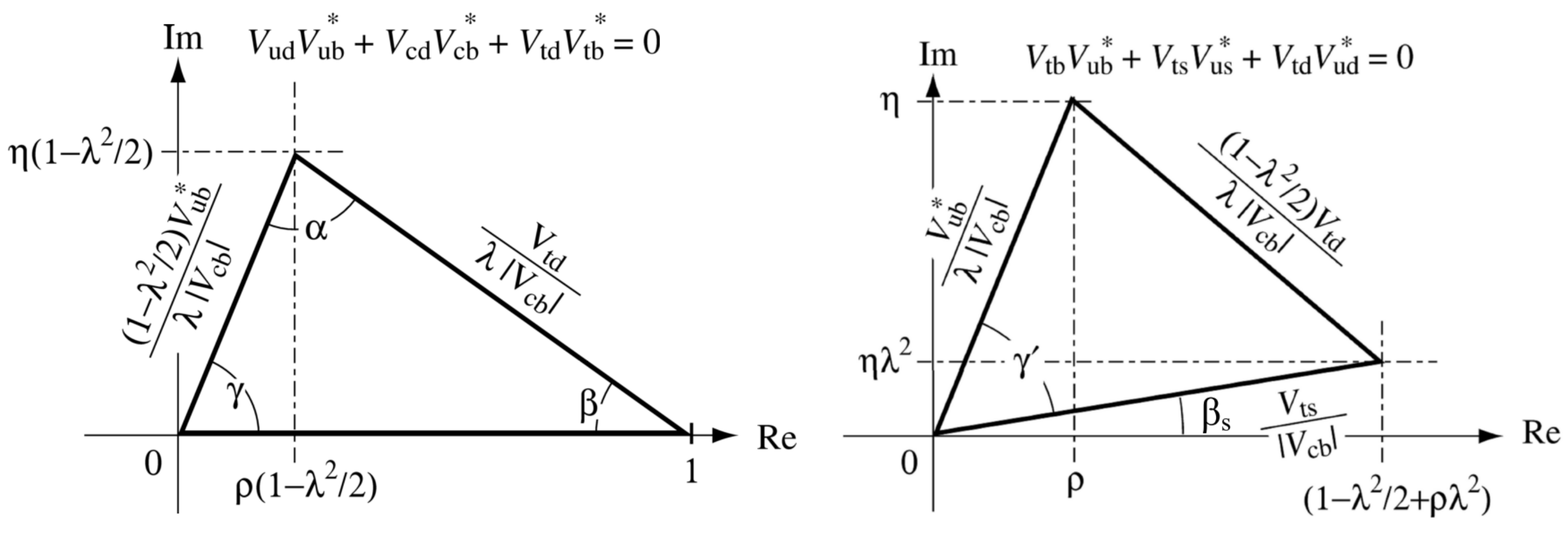}
\caption{Representation in the complex plane of the nonsquashed
triangles obtained from the off-diagonal unitarity relations of
the CKM matrix (Equation~\ref{eq:nonsquashedUT}). (\textit{a}) The
three sides are rescaled by $V_{cd}V_{cb}^*$. (\textit{b})\ The three sides are scaled by
$V_{us}V^*_{cb}$.}\label{fig:unitaryTriangle}
\end{figure}
%%%%%%%%%%%%%%%%%%%%%%%%%%%%%%%%%%%%%%%%%%%%%%%%%%%%%%%%%%%%%%%%%%%
Figure~\ref{fig:unitaryTriangle} depicts these two triangles in the complex plane. 
In particular, the triangle
defined by the former equation and rescaled by a factor
$V_{cd}V_{cb}^*$ is commonly referred to as the unitarity
triangle (UT). The sides of the UT are given by
\begin{equation}\label{eq:utR}
R_{u} \equiv
\left|\frac{V_{ud}V_{ub}^{*}}{V_{cd}V_{cb}^{*}}\right| =
\sqrt{\bar{\rho}^{2}+\bar{\eta}^{2}}, \qquad R_{t} \equiv
\left|\frac{V_{td}V_{tb}^{*}}{V_{cd}V_{cb}^{*}}\right| =
\sqrt{\left(1-\bar{\rho}\right)^{2}+\bar{\eta}^{2}}.
\end{equation}
The parameters $\bar{\rho}$ and $\bar{\eta}$ are the coordinates
in the complex plane of the nontrivial apex of the UT, the others
being (0, 0) and (1, 0). \CP\  violation in the quark sector
($\bar\eta\neq0$) is translated into a nonflat UT. The angles of the
UT are related to the CKM matrix elements as

\begin{align}
\alpha&\equiv\phi_2\equiv \arg{\left(-\frac{V_{td}V_{tb}^{*}}{V_{ud}V_{ub}^{*}}\right)} = \arg{\left(-\frac{1-\bar{\rho}-i\bar{\eta}}{\bar{\rho}+i\bar{\eta}}\right)},\label{eq:alpha}\\
\beta&\equiv\phi_1\equiv\arg{\left(-\frac{V_{cd}V_{cb}^{*}}{V_{td}V_{tb}^{*}}\right)} = \arg{\left(\frac{1}{1-\bar{\rho}-i\bar{\eta}}\right)}, \label{eq:beta}\\
\gamma&\equiv\phi_3\equiv\arg{\left(-\frac{V_{ud}V_{ub}^{*}}{V_{cd}V_{cb}^{*}}\right)}
= \arg{\left(\bar{\rho}+i\bar{\eta}\right)}. \label{eq:gamma}
\end{align}
The above equations show the two coexisting notations in the literature. Because it
involves the CKM matrix $V_{Ud}V_{Ub}^*$ (where $U=u,c,t$), the UT
arises naturally in discussion of \Bd meson transitions.

The second nonsquashed triangle has similar characteristics with
respect to the UT, but it involves $V_{uD}V_{tD}^*$ (where
$D=d,s,b$) and is not immediately associated with a neutral meson. One can define a
modified triangle (Figure~\ref{fig:unitaryTriangle}) in which all sides are rescaled by $V_{us}V^*_{cb}$. Up
to $O(\lambda^4)$ corrections, its apex is located at the point
$(\rho ,~\eta)$, and it is tilted with respect to the horizontal
axis by an angle

\begin{equation}\label{eq:betas}
\beta_{s} \equiv
\arg{\left(-\frac{V_{ts}V_{tb}^{*}}{V_{cs}V_{cb}^{*}}\right)} =
\lambda^2 \bar\eta + \mathcal{O}(\lambda^{4}).
\end{equation}
As mentioned above, neutral mesons with other flavour content
(\Bs, \Dz, \Kz) would correspond to other squashed triangles
with the same area and with some of their angles related to those
defined above. For instance, $\beta_s$ occurs naturally in the \Bs
unitarity triangle defined from $V_{Us}V_{Ub}^*$ (where $U=u,c,t$).
All these representations are particular two-dimensional
projections of the four parameters describing the CKM matrix,
which can be constrained through the combination of experimental and theoretical
information.

\section{INDIVIDUAL CONSTRAINTS}\label{Sec:Constraints}

\boldsubsection{Types of Constraints}\label{Sec:Types}

Due to its economical structure in terms of only four parameters and
its consequences for \CP violation, the CKM matrix can be
determined through many different quark transitions. These
correspond to $\Delta F = 1$ decays or $\Delta F=2$ processes
related to neutral-meson mixing.

Extensive measurements have been performed on \kaon, \D, and \B
mesons at different experiments. Constraints coming from \kaon
mesons or unflavoured particles have been obtained mostly from dedicated
experiments, among which NA48~\cite{Jeitler:2002uh},
KLOE~\cite{Aloisio:1993fy,AmelinoCamelia:2010me}, and KTeV %~\cite{}
feature prominently. Measurements of CKM parameters from \D and \B
mesons were pioneered by ARGUS~\cite{Albrecht:1988vy} at DESY,
CLEO, and CLEO-c~\cite{Andrews:1982dp} at Cornell, followed by the
so-called \B factory experiments BaBar~\cite{Aubert:2001tu} at SLAC and
Belle~\cite{Abashian:2000cg} at KEK. They operated primarily at a
center-of-mass energy corresponding to the mass of the
$\PUpsilon(4S)$ resonance. Significant contributions also came
from the CDF and D0 experiments at FNAL~\cite{Kuhr:2013hd},
especially those involving \Bs mesons, which are not accessible at
the $\PUpsilon(4S)$ resonance. These experiments have been
terminated, whereas Belle is being upgraded~\cite{Abe:2010gxa}.
Physics with \bquark and \cquark hadrons is now dominated by the
LHCb experiment~\cite{Alves:2008zz} at the LHC. The
general-purpose detector experiments ATLAS~\cite{Aad:2008zzm} and
CMS~\cite{Chatrchyan:2008aa} contribute in selected areas, and
the BESIII experiment~\cite{Ablikim:2009aa} also provides many
results for charm hadrons.

A given experimental measurement is related to an amplitude that sums
several terms, each containing CKM factors multiplied by
quantities describing the quark transition and the hadronisation
of quarks into observable mesons or baryons. Whether a given
process is relevant to measurements of the CKM\ parameters depends on the experimental and theoretical
accuracy that can be reached. Due to the complexity of
long-distance strong interactions, it is easier to select
processes with a limited number of hadrons in the initial or final
state, or to select observables (typically ratios) for which uncertainties due to
long-distance QCD effects cancel.

In the first case, (exclusive) \CP-conserving processes with at
most one hadron in the initial and the final state are considered.
After heavy degrees of freedom (in particular, weak
gauge bosons) are integrated out  using the effective Hamiltonian
formalism~\cite{Buchalla:1995vs}, the long-distance hadronic
contribution can be parametrised in terms of relatively simple
quantities that are accessible through theoretical tools (lattice
QCD simulations, effective field theories): decay constants for
leptonic decays, form factors for semileptonic decays, bag
parameters (matrix elements of four-quark effective operators
between a meson and its antimeson) for neutral-meson mixing. It
is often useful to consider ratios of observables related by
$SU(3)$ flavour symmetry, as many experimental and theoretical
uncertainties decrease in such ratios. For a few (inclusive) processes, a sum over all possible final states is performed; quark--hadron duality can then be invoked to compute the effects of the strong interaction perturbatively. For this first type of observable, for which significant hadronic uncertainties must be assessed carefully, the resulting constraints are generally set on the modulus of a given CKM matrix
element, and are dominated by theoretical uncertainties.

In the second case, \CP-violating observables are devised by
comparing a process and its \CP\  conjugate. Because the strong
interaction conserves \CP, the same hadronic amplitudes are
involved and may cancel in well-designed observables such as \CP
asymmetries, measuring \CP violation in hadron decays involving
neutral-meson mixing, or in the interference between these two
types of processes. This second type of observable, from which most of the hadronic uncertainties are absent, often  yields information about one
particular angle of the UT, dominated by experimental
uncertainties. Large \CP asymmetries are associated with the
nonsquashed UT and thus occur mainly for $B$ meson processes
(often with small branching ratios due to CKM-suppressing
factors).

\begin{table}\centering
\caption{A partial list of measurements generally used to determine
the CKM parameters, the combination of CKM parameters constrained,
and the theoretical inputs needed. The measurements are classified according to the dominant type of
uncertainties (experimental or theoretical) and the type of
processes involved (tree or loop). Abbreviation:\ OPE, operator product expansion.}\label{tab:CKMconstraints}
\begin{tabular}{lllll}
\hline
& \multicolumn{2}{c}{Dominated by experimental} & \multicolumn{2}{c}{Dominated by theoretical}\\
& \multicolumn{2}{c}{uncertainties} & \multicolumn{2}{c}{uncertainties}\\
\hline
& Process & Constraint & Process & Constraint \\
\hline & & & $B\to D^{(*)}\ell\nu$ & $|V_{cb}|$ versus {form factor $F^{B\to D^{(*)}}$}\\
&& & $B\to X_c\ell\nu$ & $|V_{cb}|$ versus OPE\\
& & & $B\to \pi\ell\nu$ & $|V_{ub}|$ versus form factor $F^{B\to \pi}$\\
Tree & $B\to D^{(*)}K^{(*)}$ & $\gamma$ & $B\to X_u\ell\nu$ & $|V_{ub}|$ versus OPE\\
& & & $M\to \ell \nu $ & $|V_{UD}|$ versus decay constant $f_M$\\
& & & $ M\to N\ell\nu $ & $|V_{UD}|$ versus form factor $F^{M\to N}$\\
&&&& or $M\to N$ amplitude\\
\hline
& $B\to (c\bar{c}) K^{(*)}$ & $\beta$ & $\epsilon_K$ ($K\overline{K}$ mix) & $V_{ts}V_{td}^*$ and $V_{cs}V_{cd}^*$\\
&&&& versus bag parameter $B_K$ \\
Loop& $B\to \pi\pi,\rho\pi,\rho\rho$ & $\alpha$
& $\Delta m_d$ (\Bd\Bdb mix) & $|V_{tb}V_{td}^*|$ versus bag parameter $B_{\Bd}$\\
& $\Bs\to J/\psi \phi$ & $\beta_s$ & $\Delta m_s$ (\Bs{}\Bsb mix) & $|V_{tb}V_{ts}^*|$ versus bag parameter $B_{\Bs}$\\
\hline
\end{tabular}
\end{table}

Table~\ref{tab:CKMconstraints} summarises the processes for which a good accuracy can be reached both
experimentally and theoretically. These processes are used to assess the
validity of the Kobayashi--Maskawa mechanism for \CP violation and
to perform the metrology of the CKM parameters, assuming the
validity of the SM. Note that $\Delta F=2$
meson mixing corresponds to a flavour-changing neutral current,
and as such, it is forbidden at tree level and is only mediated by
loop processes in the SM. By contrast, $\Delta F=1$ decays
can be either related to tree processes (typically, leptonic and
semileptonic decays) or involve loop processes (such as
hadronic decays).

The potential sensitivity to physics beyond the SM is not the
same for all processes: When discussing potential NP effects, it
is often interesting to perform the metrology of the CKM
matrix using only tree-level processes (this is possible by use of the unitarity of the CKM matrix and the fact that CKM moduli, apart from $V_{td}$ and $V_{ts}$, can all be measured from tree-level decays) and to exploit loop processes in order to constrain additional NP effects. One may also consider additional
ultrarare decays and processes that are not experimentally measured with sufficient accuracy to constrain the CKM matrix in the SM,
but are very sensitive to NP---for instance the rare $\Bs\to\mu\mu$
and $K\to \pi\nu\nu$ decays or the \Bs width difference
$\Delta\Gamma_s$. This issue is discussed further in
Section~\ref{Sec:globalNP}.

\boldsubsection{Moduli from Leptonic and Semileptonic Decays $\Delta F = 1$}\label{Sec:Moduli}

The moduli described in the following subsections can be determined accurately from
(\CP-averaged) branching ratios of exclusive leptonic and
semileptonic decays.

\boldsubsubsection{Transitions among the first and second generations}

The CKM matrix element $|V_{us}|$ is efficiently constrained by
\decay{\Km}{\ellm\neub}, \decay{K}{\pi\ell\neub} and
\decay{\tau}{\Kz\nu_\tau} decays~\cite{PDG2016}. Decay constants
and form factors are known from lattice QCD
simulations~\cite{Aoki:2016frl}, whereas radiative corrections have
been determined with a high accuracy on the basis of chiral perturbation
theory~\cite{Antonelli:2010yf}.

The matrix elements $|V_{cd}|$ and $|V_{cs}|$ are constrained by
\D, \Dp, and \Ds leptonic and semileptonic decays. The precision
of the leptonic
decays~\cite{Ablikim:2013uvu,Zupanc:2013byn,Alexander:2009ux,delAmoSanchez:2010jg}
[where the lepton is often a muon but can be a $\tau$ lepton in
the case of the \Ds
meson~\cite{delAmoSanchez:2010jg,Onyisi:2009th,Naik:2009tk}] is
dominated by experimental uncertainties. Conversely, the
semileptonic \decay{D}{K\ell\nu} and \decay{D}{\pi\ell\nu}
decays~\cite{Ablikim:2015ixa,Lees:2014ihu,Besson:2009uv,Widhalm:2006wz,Aubert:2007wg}
have not been investigated by many lattice QCD collaborations, and
their systematic uncertainties are expected to be improved to
yield relevant constraints for the CKM
parameters~\cite{Aoki:2016frl}. Moreover, radiative corrections
still need to be investigated in detail for these
processes~\cite{Burdman:1994ip,Becirevic:2009aq}.

In principle, $|V_{ud}|$ could be determined by many processes, such as
$\pi^+\to e^+\nu$, $\pi^+\to\pi^0 e^+\nu$, and $n\to p e^-\bar\nu$.
Yet they exhibit poor experimental accuracy for our purposes (pion leptonic or semileptonic decays), or their measurements in different experimental settings are not compatible and cannot be averaged meaningfully (neutron lifetime)~\cite{PDG2016}. It turns out that the
most accurate determination comes from nuclear superallowed
$0^+\to 0^+$ $\beta$ decays~\cite{Hardy:2014qxa}. The current
determination is based on a large set of nuclei and relies on
sophisticated estimates of different corrections (electroweak
radiative, nuclear structure, isospin violation) from dedicated
nuclear physics approaches.

\boldsubsubsection{$|\Vub|$ and $|\Vcb|$}

The determination of the CKM matrix elements $|\Vub|$ and $|\Vcb|$
provides important closure tests of the UT. It is best performed
in semileptonic \decay{\bquark}{(\uquark,\cquark)\ell\nu} decays
$(\ell=e,\mu)$, where there are no hadronic uncertainties related
to the decay of the emitted \W boson. Unfortunately, a well-known
discrepancy exists between the determinations obtained from
exclusive decays and from inclusive modes~\cite{HFAG}, which are
treated with different tools. In the case of \Vcb, there is no
complete lattice QCD determination of the $B\to D^{(*)}\ell\nu$
form factors, which are required in order to analyse the corresponding
experimental exclusive
measurements~\cite{Aubert:2008yv,Dungel:2010uk,Aubert:2009ac,HFAG}.
Heavy-quark effective theory (HQET) is required, expanding the
form factors in powers of $1/m_b$ and $1/m_c$ in order to simplify
their expression and constrain their dependence on the lepton
energy, complemented with lattice QCD estimates of some of the
HQET parameters. For the inclusive decay $B\to X_c\ell\nu$~\cite{Urquijo:2006wd,Schwanda:2006nf,Aubert:2004td,Aubert:2009qda},
operator product expansion (OPE) \cite{Shifman:1978bx,Shifman:1978by} allows the decay rate to be expressed as a
series in $1/m_b$ and $1/m_c$~\cite{Alberti:2014yda}, with matrix
elements that can be fitted from leptonic and hadronic moments of
the branching ratio~\cite{Gambino:2013rza}.

In the case of $|\Vub|$, the exclusive determination benefits from
lattice QCD computations for the vector form factors of the decay
$B\to \pi\ell\nu$~\cite{Lattice:2015tia,Flynn:2015mha,Dalgic:2006dt}, which
can be combined with measurements of the differential decay
rate~\cite{HFAG,Ha:2010rf,Lees:2012vv,delAmoSanchez:2010af} in
order to determine \Vub. The inclusive
determination~\cite{Lees:2011fv,Aubert:2005mg,Aubert:2005im,Limosani:2005pi,Bornheim:2002du}
is more challenging. The full decay rate cannot be accessed, because
a cut in the lepton energy must be performed to eliminate the huge
$b\to c\ell\nu$ background. The OPE must be
modified, introducing poorly known shape functions describing the
$b$ quark dynamics in the $B$
meson~\cite{Lange:2005yw,Bauer:2001rc,Aglietti:2007ik,Gambino:2007rp,Gardi:2008bb,Bigi:1993fe,Bigi:1993ex}.
They can be constrained partly from \decay{B}{X_s\gamma} and raise
questions concerning the convergence rate of the series in
$1/m_b$~\cite{Schwanda:2008kw}.

\begin{figure}\centering
\includegraphics[width=\textwidth]{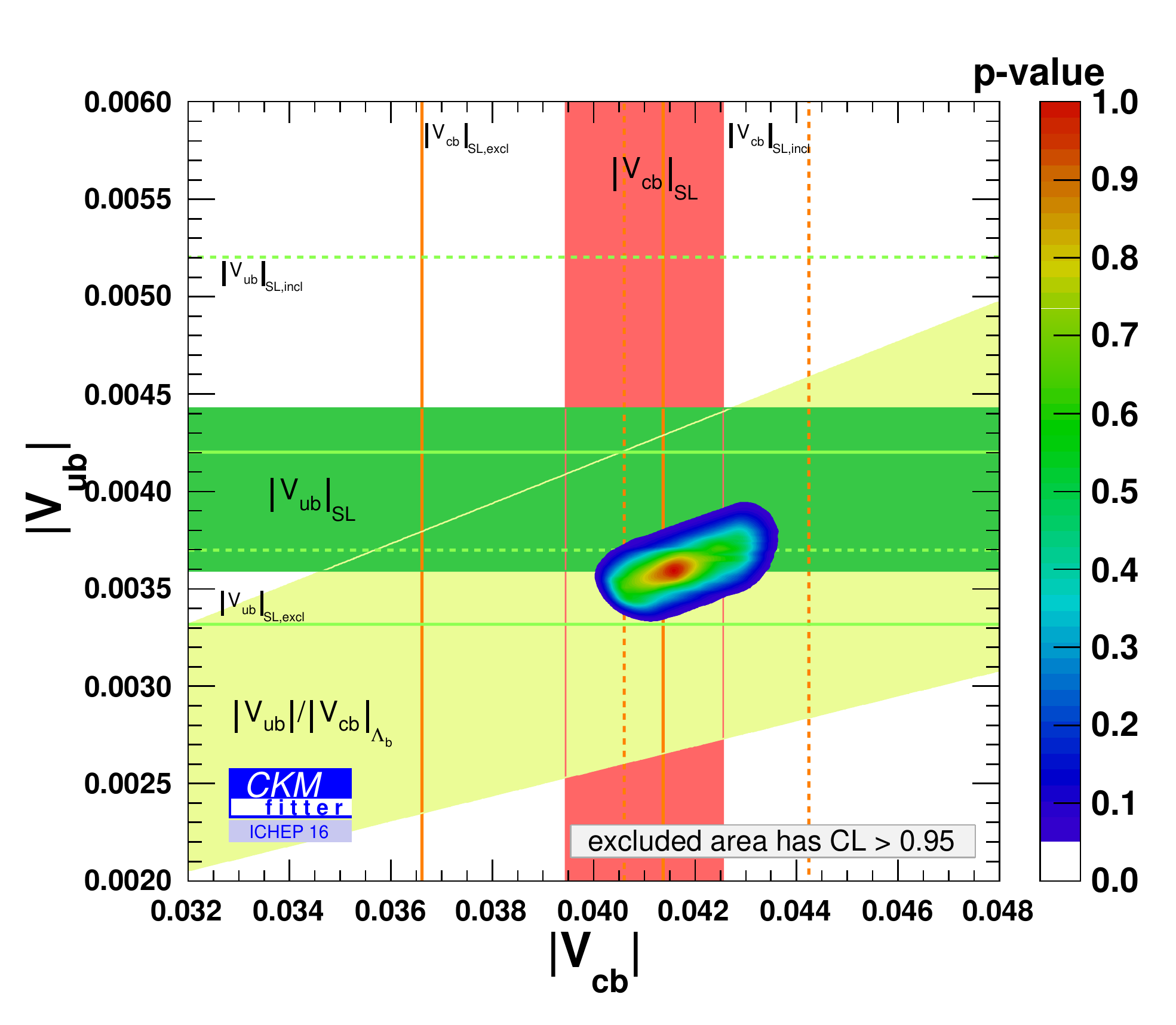}
\caption{Experimental situation for $|\Vub|$ and $|\Vcb|$. The
experimental measurements from exclusive (inclusive) measurements
are represented by bands with solid (dashed) lines, and their
average is represented by the coloured bands. The yellow diagonal
band corresponds to the constraint from \Lb decays. The oval
region indicates the 95\%-CL region is the indirect determination
of $|\Vub|$ and $|\Vcb|$ from a global fit including none of these
measurements~\cite{Charles:2004jd}.}\label{Fig:2015-013}
\end{figure}

These determinations leads to a long-standing discrepancy between
inclusive and exclusive determinations for $|\Vub|$ and $|\Vcb|$.
Currently, global fits (discussed in
Section~\ref{Sec:global}) use averages of both kinds of
determination as inputs, but their outcome favours exclusive
measurements for $|\Vub|$ and inclusive measurements for $|\Vcb|$
(Figure~\ref{Fig:2015-013}).

Additional decay modes need to be added in order to obtain a global picture
for $|\Vub|$ and $|\Vcb|$. The leptonic decay $B\to\tau\nu_\tau$
has been studied at \B
factories~\cite{Lees:2012ju,Kronenbitter:2015kls,Adachi:2012mm,Aubert:2009wt},
favouring values closer to the inclusive determination. The value
of this branching ratio used to be at odds with expectations from
global fits~\cite{Lenz:2010gu}, but recent determinations from
Belle reduced the discrepancy to 1.2$\sigma$. In addition,
the LHCb Collaboration recently used \Lb baryon decays for the
first time \cite{LHCb-PAPER-2015-013}. The decay rates of
\decay{\Lb}{\proton\mun\nu} and \decay{\Lb}{\Lc\mun\nu} are
compared to determine the ratio $|\Vub/\Vcb|$, using the available
lattice QCD estimates of the six different form factors
involved~\cite{Detmold:2015aaa}. Figure~\ref{Fig:2015-013} depicts the overall situation, including the constraints from inclusive
and exclusive determinations of $|\Vub|$, $|\Vcb|$, and
$|\Vub/\Vcb|$.

\boldsubsubsection{$|V_{tb}|$, $|V_{td}|$, and $|V_{ts}|$}

The measurement of $|V_{tb}|$ can be performed from the cross section for single top quark production. The combination of Tevatron and LHC data yields $|V_{tb}|=1.021\pm 0.032$~\cite{PDG2016}, which is not competitive within the SM with the very accurate determination based on unitarity and other constraints on the CKM parameters. Other, less stringent constraints on $|V_{tb}|$ can be obtained from the ratio of branching ratios ${\rm Br}(t\to W b)/{\rm Br}(t\to W q)$ and from LEP electroweak precision measurements. In principle, the matrix elements $|V_{td}|$ and $|V_{ts}|$ can be measured directly from tree-level decays of top quarks~\cite{Lacker:2012ek}, but the results are not competitive with neutral-meson mixing within the SM (see Section 3.4).

\boldsubsection{Unitary Triangle Angles from $CP$-Violating Measurements}\label{Sec:CPV}
The UT angles described in the following subsections can be determined experimentally from $CP$-violating measurements with almost no theoretical uncertainties.
\boldsubsubsection{The angle $\beta\equiv\varphi_1$}

The mode that allowed for the first observation of \CP violation
in \B decays is
\decay{\Bd}{\jpsi\KS}~\cite{Aubert:2001nu,Abe:2001xe}. It provides
access to $\varphi_d$~\cite{Carter:1980tk}, the relative phase
between the decay of the \Bz meson to {\jpsi\KS} and that of the
oscillation of \Bd to its antiparticle \Bdb, followed by the decay
\decay{\Bzb}{\jpsi\KS}. The measurement requires studying how the
decay depends on the time between the initial production of \Bd
and its decay, leaving time for evolution and potential mixing
between \Bd and \Bdb mesons. In the SM, the decay is dominated by
a single CKM phase, up to Cabibbo-suppressed penguin
contributions, whereas  \Bd mixing is completely dominated by
top--top box diagrams. Considering these two amplitudes, the
measurement of the time dependence of this process yields
$\sin2\beta$~\cite{Bigi:1983cj,Bigi:1981qs,Carter:1980tk}. The \B factories were optimised
for this
measurement~\cite{Bevan:2014iga,Aubert:2009aw,Adachi:2012et} and
determined~\cite{HFAG} $\sin2\beta^\text{B-fact} = 0.682 \pm
0.019$, which is the most precise constraint on the UT (Figure~\ref{Fig:CKMFit1}). Recently, LHCb joined the
effort, publishing its first measurement of the time-dependent
\CP asymmetry in the decay \decay{\Bd}{\jpsi\KS}
\cite{LHCb-PAPER-2015-004} with an uncertainty competitive with
the individual measurements from the \B factories. The
degeneracies among the values of $\beta$ are lifted thanks to the
$\Bd\to J/\psi\Kstarz$ mode~\cite{Aubert:2004cp,Itoh:2005ks},
where the interferences between the difference partial waves are
sensitive to $\cos 2\beta$.

The measured value for $\sin 2\beta$ is slightly lower than the
expectation from all other constraints on the
UT~(\cite{Charles:2015gya}), $\sin2\beta^\text{indirect} =
0.740^{+0.020}_{-0.025}$, which could be due to the so-far-neglected contribution from penguin topologies in the decay
\decay{\Bd}{\jpsi\KS} or in other
\decay{\bquark}{\cquark\cquarkbar\squark} decays to \CP
eigenstates. There have been several theoretical attempts to
estimate this contribution. One possibility consists of using
$SU(3)$ symmetry and assessing the size of penguin contributions
from \decay{\Bd}{\jpsi\piz}, \decay{\Bd}{\jpsi\rhoz}, and
\decay{\Bs}{\jpsi\KS}
decays~\cite{DeBruyn:2014oga,DeBruyn:2010hh}; unfortunately, the accuracy of any constraints from these studies is
currently limited due to the experimental
inputs \cite{Aubert:2008bs,Lee:2007wd,LHCb-PAPER-2014-058,LHCb-PAPER-2015-005}.
By contrast, a fit to $B\to J/\psi P$ (where $P$ is a light pseudoscalar meson)
including $SU(3)$ breaking corrections suggests a small contamination from penguin
contributions~\cite{Jung:2012mp}. Direct computations based on
soft-collinear effective theory arguments~\cite{Frings:2015eva}
reach a similar conclusion. The final average of all charmonium
data yields the very accurate value $\sin2\beta^\text{meas} =
0.691 \pm 0.017$~\cite{HFAG}.

The value of $\sin 2\beta$ can also be determined in
\decay{\bquark}{\quark\quarkbar\squark} transitions (where $q=d,s$)
as \decay{\Bd}{\eta'\Kz}~\cite{Santelj:2014sja,Aubert:2008ad}.
These transitions are not allowed at tree level and thus probe the CKM
mechanism in loop-induced processes, although contamination
from penguins with other CKM phases is difficult to assess in
these modes~\cite{Beneke:2005pu}. The na\"ive average of all
measurements results in $\sin2\beta^{\quark\quarkbar\squark}=0.655
\pm 0.032$~\cite{HFAG}, which is consistent with expectations.

Crucial to this and other time-dependent measurements is
the ability to identify the flavour of the \B meson, before it
starts its evolution and mixes with its antiparticle. Whereas so-called flavour tagging had a high
efficiency at \B
factories \cite{Bevan:2014iga}, at the LHC the complicated hadronic
environment makes this task very challenging. The
tagging performance at LHCb has continuously improved over the
years thanks to both a better understanding of the underlying event and
the use of modern machine earning
techniques~\cite{LHCb-PAPER-2011-027,LHCb-PAPER-2015-027,LHCb-PAPER-2015-056,LHCb-PAPER-2016-039}.
These improvements, in combination with data from the upcoming LHC Run 2,
will enable further reduced uncertainties.

\boldsubsubsection{The angle $\alpha\equiv\varphi_2$}

A precise determination of the UT angle $\alpha$ is challenging
at both the theoretical and experimental levels. It requires the
time-dependent study of \decay{\bquark}{\uquark} transitions as in
$B\to \pi\pi$, $B\to \rho\pi$, or $B\to\rho\rho$, which are
affected by \decay{\bquark}{\dquark} or \decay{\bquark}{\squark}
penguin topologies, depending on the final state considered. The
interference between \Bd--\Bdb mixing and decay amplitudes would
provide a measurement of $\pi-\beta-\gamma=\alpha$ (using
unitarity) in the absence of penguin contributions. In practice,
this penguin pollution is present and must be constrained by
determining the magnitude and the relative phases of hadronic
amplitudes before determining the angle $\alpha$, with the help of
isospin symmetry~\cite{Gronau:1990ka,Lipkin:1991st}. For
$B\to\pi\pi$~\cite{Lees:2012mma,Adachi:2013mae,LHCb-PAPER-2013-040,Bornheim:2003bv,Aubert:2007hh,Duh:2012ie,Abe:2004mp},
all three possible channels are considered, and isospin symmetry
can be used to relate the hadronic amplitudes, leading to
triangular relations. From the measurements of branching ratios
and \CP asymmetries, two triangles can be reconstructed for $B^+,
\Bd$ and $B^-, \Bdb$ decays, respectively, with a relative angle
corresponding to $\alpha$, up to discrete ambiguities. For the decay  $B\to
\rho\rho$~\cite{Aubert:2007nua,Vanhoefer:2015ijw,Aubert:2009it,Zhang:2003up,Aubert:2003wr,Kusaka:2007mj,Jessop:2000bv},
a similar construction can be invoked for the (dominant)
longitudinal polarisation; interestingly, the
penguin contamination turns out to be less important than for
$\pi\pi$ modes. The decay $B\to
\rho\pi$~\cite{Aubert:2007py,Zhang:2004wza,Kusaka:2007mj,Aubert:2003fm,Jessop:2000bv}
requires a more elaborate analysis: Isospin symmetry yields
pentagonal relations, whereas the time-dependent $B\to\pi\pi\pi$
Dalitz plot analysis provides a large set of observables,
corresponding to the parametrisation of the amplitude together
with an isobar model involving the $\rho$ line shape. So far, a Dalitz plot analysis has been reported only for the decay mode \decay{\Bp}{\pip\pim\pip} decay mode ~\cite{Aubert:2009av}. The present average of these
constraints yields $\alpha^{\rm
meas}=(88.8^{+2.3}_{-2.3})\degrees$~\cite{Charles:2017evz}.

\begin{figure}\centering
\includegraphics[width=\textwidth]{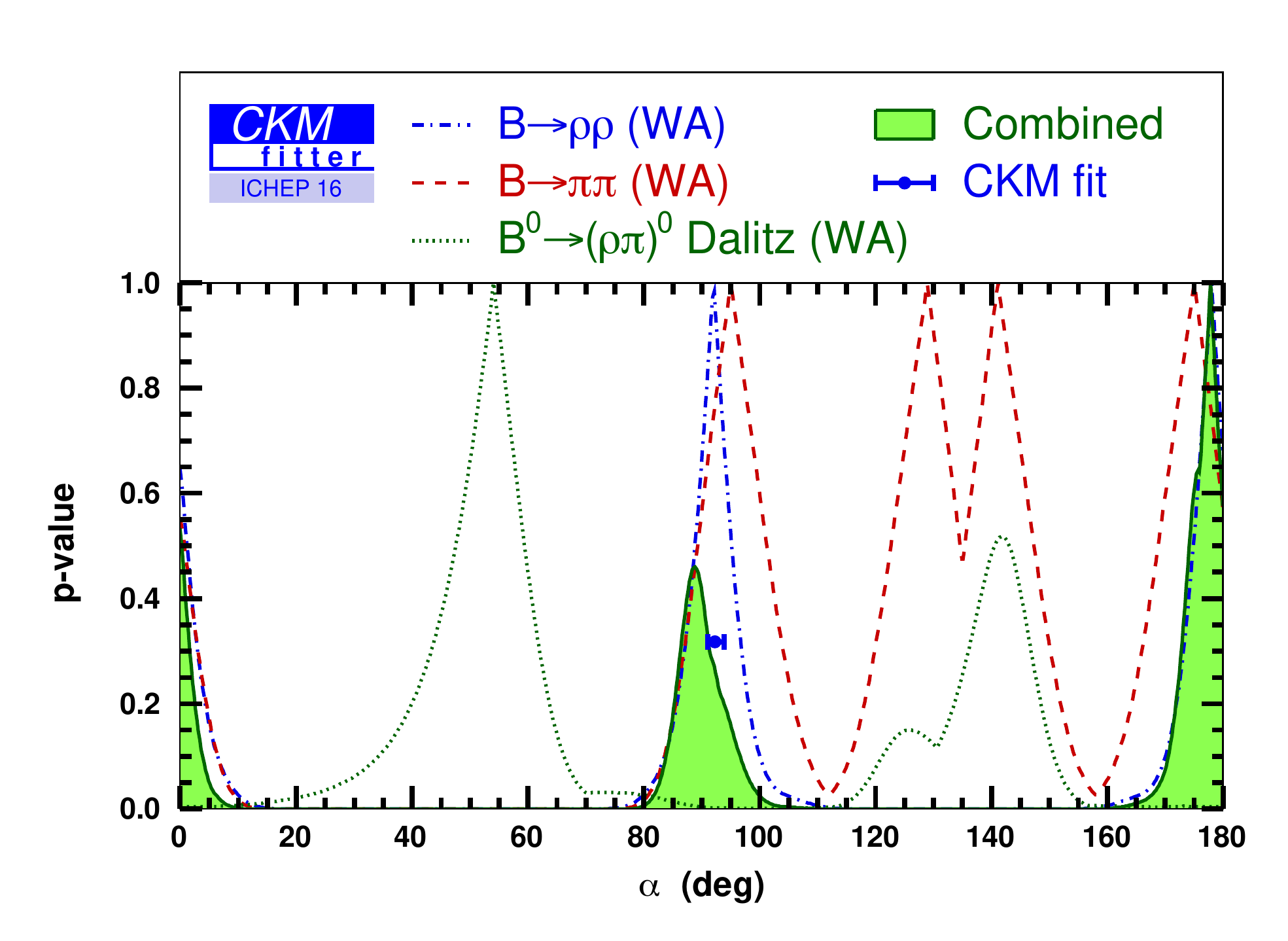}
\caption{Constraints on the CKM angle $\alpha$ from $B\to \pi\pi$,
$B\to \rho\pi$, and $B\to \rho\rho$. The combination of the
constraints and the outcome of the global fit are also
represented~\cite{Charles:2017evz}. }\label{Fig:alpha}
\end{figure}

Figure~\ref{Fig:alpha} depicts the different constraints and shows the discrete symmetries present in
the $\pi\pi$ and $\rho\rho$ cases, as well as the fact that two
solutions are allowed by the combination of the measurements. In
addition to the statistical uncertainties of the measurements, the
accuracy is limited by two main hypotheses: $\Delta I=3/2$
contributions coming from electroweak penguins are neglected, and
isospin symmetry in strong interactions is not
perfect~\cite{Gronau:2005pq,Zupan:2007fq,Charles:2017evz}.

\boldsubsubsection{The angle $\gamma\equiv\varphi_3$}\label{Sec:gamma}

The angle $\gamma$ can be obtained from tree-dominated
\decay{\B}{\D\kaon} decays, where the \CP-violating phase appears
in the interference of \decay{\bquark}{\cquark} (colour-allowed)
and \decay{\bquark}{\uquark} (colour-suppressed) topologies,
followed by carefully chosen $D$ decay processes. This is the least
precisely known angle of the UT, and its
determination from tree decays is considered to be free from
contributions beyond the SM and unaffected by hadronic
uncertainties, contrary to the angles $\alpha$ and
$\beta$~\cite{Brod:2013sga}. There is still great potential for the improvement of the measurement of $\gamma$ by several order of magnitudes compared with the theoretical uncertainties. The angle $\gamma$ can thus provide a reference to which other measurements of the CKM parameters can be compared both within the SM and beyond.

Three different methods have been devised in order to obtain
information on $\gamma$, depending on the subsequent decays of
$D^{(*)}$ mesons, with a different sensitivity to the ratio of
colour-favoured and colour-suppressed amplitudes. The
Gronau--London--Wyler (GLW) method~\cite{Gronau:1990ra,Gronau:1991dp}  considers the decay
of the \D meson into \CP eigenstates, eliminating further hadronic
uncertainties concerning the \D decays. The Atwood--Dunietz--Soni (ADS)
method~\cite{Atwood:1996ci,Atwood:2000ck} considers decays of the
$D^{(*)}$ meson with a pattern of Cabibbo dominance/suppression
that counteracts the colour suppression/dominance of the $B$
decay, for instance, $D\to \Kmp\pipm$. Finally, the Giri--Grossman--Soffer--Zupan (GGSZ) method~\cite{Giri:2003ty} performs a Dalitz analysis of three-body
$D^{(*)}$ decays, inducing a dependence on the amplitude model for
$D^{(*)}$ decays.

The last two methods require additional information about the strong
phase structure in multibody \D decays, which was
provided by
CLEO-c~\cite{Insler:2012pm,Malde:2015mha,Evans:2016tlp}. LHCb has
performed several measurements using the
GLW/ADS~\cite{LHCb-PAPER-2016-003,LHCb-PAPER-2015-014,LHCb-PAPER-2015-020,LHCb-PAPER-2015-059,LHCb-PAPER-2013-068}
and GGSZ~\cite{LHCb-PAPER-2014-041,LHCb-PAPER-2016-007} methods
with various \Bd and \Bp decays, as well as a time-dependent
\decay{\Bs}{\Dspm\Kmp}
analysis~\cite{LHCb-PAPER-2014-038,LHCb-CONF-2016-015}. As some
systematic uncertainties are correlated among analyses, LHCb has
performed a combination yielding
$\gamma=(72.2^{+6.8}_{-7.3})\degrees$~\cite{LHCb-PAPER-2016-032}.

\begin{figure}\centering
\includegraphics[width=\textwidth]{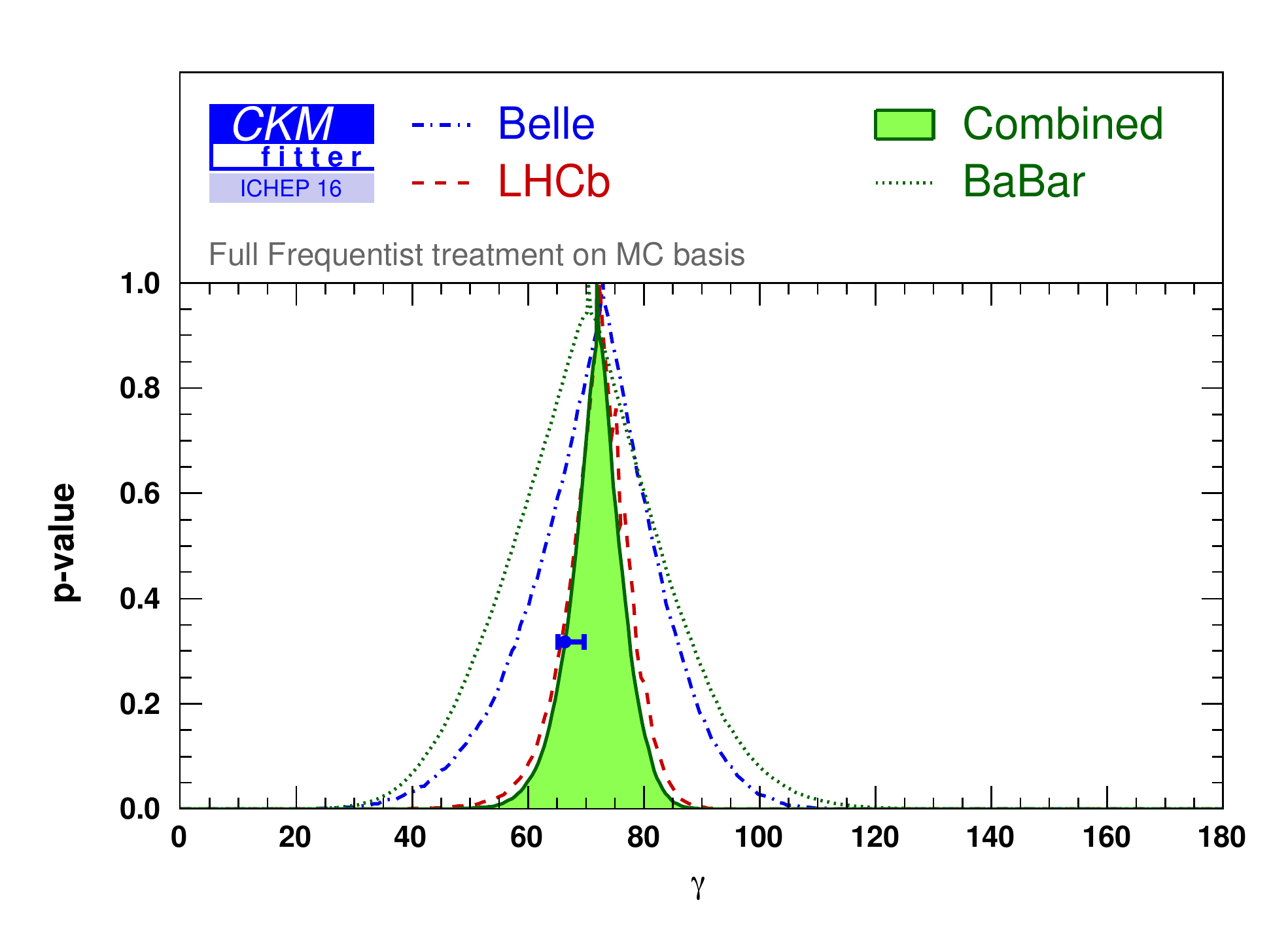}
\caption{Constraints on the CKM angle $\gamma$ from the BaBar, Belle,
and LHCb experiments~\cite{Charles:2004jd}.}\label{Fig:gamma}
\end{figure}

Similarly, the \B factories BaBar and Belle have performed
combinations of their
measurements~\cite{Lees:2013nha,Trabelsi:2013uj} and obtained
$\gamma=(67\pm11)\degrees$~\cite{Bevan:2014iga}. The combination
of the values for $\gamma$ yields $\gamma^{\rm
meas}=(72.1^{+5.4}_{-5.8})^\circ$, with the confidence-level
curves shown in Figure~\ref{Fig:gamma}. Because there is no irreducible
theoretical uncertainty on the determination of
$\gamma$~\cite{Brod:2013sga}, there is plenty of room for more
precision measurements of this quantity.

\boldsubsubsection{The angle $\varphi_s$}\label{Sec:phis}

\begin{figure}\centering
\includegraphics[width=\textwidth]{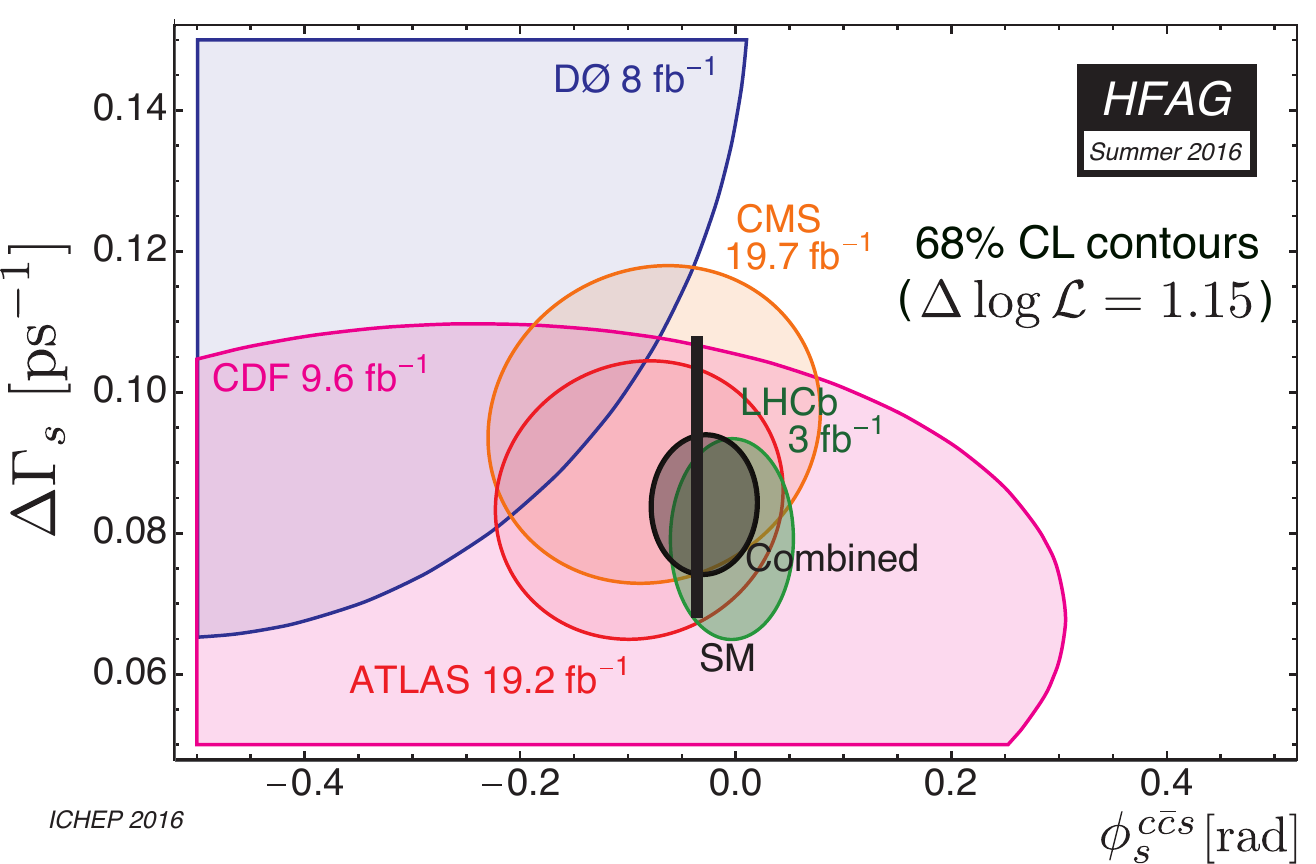}
\caption{Constraints on $\Delta\Gamma_s$ and
$\varphi_s^{\cquark\cquarkbar\squark}$ from various decays and
experiments~\cite{HFAG}. The Standard Model (SM)
predictions are from
References~\cite{Lenz:2011ti,Lenz:2006hd,Charles:2004jd}.}
\label{Fig:2014-059}
\end{figure}

By analogy with the measurement of $\sin 2\beta$ related to \Bd
mixing, a \CP-violating phase $\varphi_s$ related to \Bs mixing
can be determined through time-dependent measurements of $b\to
c\bar{s}s$ decays. This phase is equal to
$-2\beta_s\equiv-2\arg[-\Vts\Vtbs/\Vcs\Vcbs]=-0.0370\aerr{0.0006}{0.0007}$
rad in the SM~\cite{Charles:2004jd}, neglecting subleading penguin
contributions. This phase has been measured using the decay \decay{\Bs}{\jpsi\phi} with \decay{\jpsi}{\mumu} and
\decay{\phi}{\Kp\Km} by CDF~\cite{Aaltonen:2012ie},
\dzero~\cite{Abazov:2011ry}, CMS~\cite{Khachatryan:2015nza}, and
ATLAS~\cite{Aad:2016tdj}. LHCb uses the decay
\decay{\Bs}{\jpsi\Kp\Km} (including \decay{\Bs}{\jpsi\phi}) in a
polarisation-dependent way~\cite{LHCb-PAPER-2014-059}, as well as the pure
\CP-odd decay
\decay{\Bs}{\jpsi\pip\pim}~\cite{LHCb-PAPER-2014-012,LHCb-PAPER-2014-019}.
Figure~\ref{Fig:2014-059} shows the current constraints on $\varphi_s$ and the decay width
difference $\Delta\Gamma_s=\Gamma_L-\Gamma_H$.

Similarly to the case for the angle $\beta$, the SM prediction
$\varphi_s^{\cquark\cquarkbar\squark}=-2\beta_s$ assumes
tree-dominated decays. With the increasing precision on the CKM
parameters, the effects of suppressed penguin topologies can no longer be
neglected~\cite{Fleischer:1999nz,Fleischer:1999zi,Fleischer:1999sj,Faller:2008zc,DeBruyn:2010hh,Ciuchini:2005mg}.
Cabibbo-suppressed decay modes, in which these topologies are
relatively more prominent, can be used to constrain such effects.
Methods of using selected measurements constraining the sizes of penguin
amplitudes have been described elsewhere~\cite{DeBruyn:2014oga,DeBruyn:2048174,Ligeti:2015yma,Jung:2012mp,Frings:2015eva}.
The LHCb Collaboration is pursuing this program with studies of
the decays \decay{\Bs}{\jpsi\KS}~\cite{LHCb-PAPER-2013-015} and
\decay{\Bs}{\jpsi\Kstarzb}~\cite{LHCb-PAPER-2015-034}.

Another interesting test of the SM is provided by the measurement
of the mixing phase $\varphi_s^{\squark\squarkbar\squark}$ with the
penguin-dominated mode \decay{\Bs}{\phi\phi}. In this case, the
measured value is $-0.17\pm 0.15\pm
0.03$~rad~\cite{LHCb-PAPER-2014-026}, which is compatible with the
SM expectation.

\boldsubsection{Information from $\Delta F=2$ Transitions}\label{Sec:DeltaF2}

$\Delta F=2$ transitions are particularly useful both in the SM
and in the search for NP, as these are flavour-changing neutral
currents arising only as loops in the SM. Among the four neutral
mesons available, the \Kz, \Bd, and \Bs systems are useful for the
metrology of the SM. Indeed, the mixing of the charm meson \Dz is
notoriously difficult to estimate theoretically because, due to the Glashow--Iliopoulos--Maiani (GIM)
mechanism, it is dominated by the first two
generations, and thus by long-distance QCD
dynamics~\cite{Falk:2001hx}.

\boldsubsubsection{\Bd and \Bs systems}\label{Sec:DeltaF2BdBs}

Due to neutral-meson oscillation, the flavour eigenstates $P^0$ and
$\bar{P}^0$ mix into the mass eigenstates $P_L$ and $P_H$, denoting
respectively, the light and heavy mesons. This language is used to
describe several observables for the \Bd and \Bs systems: the mass
difference $\Delta m=M_H-M_L$, the width difference $\Delta
\Gamma=\Gamma_L-\Gamma_H$, and the semileptonic asymmetry
$a_\text{SL}^{d,s}$ that measures \CP violation in mixing by comparing
semileptonic decays of $P^0$ or $\bar{P}^0$ into ``wrong-sign''
leptons (such processes can occur only if $P_0$ or $\bar{P}_0$ mixes into its antiparticle). Due to the pattern of
CKM factors (suppressing charm contributions), $\Delta m$ is
dominated by the dispersive part of top quark--dominated box diagrams. It can be studied
within an effective Hamiltonian analysis by integrating out heavy
$(W,Z,t,H)$ degrees of freedom: It amounts to a local contribution
that requires the input of a single bag parameter once
short-distance QCD corrections (gluon exchanges) have been taken
into account~\cite{Buchalla:1995vs,Nierste:2009wg}. This explains
why the mass difference $\Delta m$ has long been used to constrain the CKM parameters. By contrast, $\Delta\Gamma$,
related to the imaginary part of the amplitude, involves only real
intermediate states. Therefore, it is dominated by the absorptive part
of the box diagram involving the charm quark, namely the decays of $P^0$ and $\bar{P}^0$ into
common final states. The evaluation of this nonlocal contribution
requires a further $1/m_b$ expansion, with larger uncertainties
and two hadronic bag parameters, making $\Delta\Gamma$ (and
$a_\text{SL}^{d,s}$) harder to control
theoretically~\cite{Beneke:1996gn,Beneke:1998sy,Ciuchini:2003ww,Lenz:2006hd,Lenz:2010gu,Bigi:1992su}.

The frequency of \Bd and \Bs mixing probes $|V_{tb}V_{tq}^\ast|$,
where $q=d$ and $s$, respectively. They are measured as $\Delta m_d^{\rm
meas}=506.4 \pm 1.9\invns$ and $\Delta m_s^{\rm meas}=17.757 \pm
0.021\invps$~\cite{HFAG}, placing strong constraints on the
UT. The accuracy of these constraints is limited mainly by the
determination of the corresponding bag parameters. It is more
useful to consider the ratio $\Delta m_d/\Delta m_s$, which
involves an $SU(3)$ breaking ratio of bag parameters that is known more
accurately than individual quantities from lattice
simulations~\cite{Aoki:2016frl}.

The \Bs meson system has many features in common with the
\Kz meson system, with a heavy long-lived and a light short-lived
eigenstate. The a priori unknown admixture of the two states
contributing to a given non-flavour-specific decay causes
uncertainties in the measurement of branching fractions, for
instance, for the decay
\decay{\Bs}{\mumu}~\cite{DeBruyn:2012wj,DeBruyn:2012wk,LHCb-PAPER-2014-049,Aaboud:2016ire}.
Thus, a precise determination of the decay width difference is also
important for the study of rare decays and efficiently~constrains models of NP
in $\Delta F=2$ transitions
\cite{Beneke:1998sy,Beneke:2002rj,Lenz:2006hd,Lenz:2010gu,Lenz:2012az}.

Whereas measurements of $\Delta m_d$ and $\Delta m_s$ are consistent
with expectations, the \dzero experiment reported an unexpectedly
large dimuon asymmetry~\cite{Abazov:2013uma} that differs from
the SM expectation by $3\sigma$. This measurement is generally
interpreted as a combination of the semileptonic asymmetries
$a_\text{SL}^d$ and $a_\text{SL}^s$ in \Bd and \Bs decays,
respectively, which measure \CP violation in mixing. Direct
measurements of $a_\text{SL}^d$ and $a_\text{SL}^s$ at \B
factories~\cite{Lees:2014kep,Lees:2013sua,Nakano:2005jb},
\dzero~\cite{Abazov:2012zz,Abazov:2012hha}, and
LHCb~\cite{LHCb-PAPER-2014-053,LHCb-PAPER-2016-013} are consistent
with the SM prediction and in tension with the \dzero asymmetry.
The origin of this discrepancy is still under
investigation~\cite{Borissov:2013wwa}, as we discuss further in
Section~\ref{Sec:DeltaF2NP}.

\boldsubsubsection{The \Kz system}

The pattern of CKM factors requires loops involving top and
charm quarks to be considered in the case of the kaon system.
The mass difference $\Delta m_K$ thus gets not only top box
contributions but also charm--top and charm--charm contributions,
which are long-distance contributions that are difficult to
estimate~\cite{Buchalla:1995vs,Nierste:2009wg}. A way out involves considering observables related to \CP violation in \Kz mixing
and decays into pions. In the absence of \CP violation, only the
short-lived kaon, $\KS$, decays into $\pi\pi$, whereas the
long-lived kaon, $\KL$, decays into $3\pi$. A measurement of \CP
violation can be defined from the amplitude of \KS and \KL states
decaying into a $\pi\pi$ state with total isospin $I=0$:

\begin{equation}
\epsilon_K=\frac{\langle (\pi\pi)_{I=0}|\KL\rangle}{\langle
(\pi\pi)_{I=0}|\KS\rangle}.
\end{equation}
This term is related to the difference between \CP eigenstates and
mass eigenstates, and it requires a global fit to many observables
describing $K\to 2\pi$ decays~\cite{PDG2016}. Its real part
indicates \CP violation in mixing, and its imaginary part measures
\CP violation in the interference between mixing and decay.  $\epsilon_K$ can be computed accurately in terms of
short-distance (Inami--Lim) functions as well as a long-distance
bag parameter, which is known from lattice QCD
simulations~\cite{Aoki:2016frl}. An accurate SM prediction of
$\epsilon_K$ also requires a resummation of short-distance QCD
corrections (gluon exchanges), encoded into
$\eta_{tt},\eta_{ct}$, and $\eta_{cc}$. These coefficients have been
computed up to next-to-leading order (NLO) for $\eta_{tt}$~\cite{Buras:1990fn} and next-to-next-to-leading order (NNLO)
for $\eta_{ct}$ and $\eta_{cc}$~\cite{Brod:2010mj,Brod:2011ty}, the latter of which is still affected by large theoretical uncertainties.
The interpretation in terms of the CKM parameters involves $A$,
$\bar\rho$, and $\bar\eta$ (and is thus connected with $|V_{cb}|$)
and corresponds to a hyperbola in the $(\bar\rho,\bar\eta)$ plane.

Another interesting quantity is given by $\epsilon'_K$, which is defined to
measure \CP violation in decays by comparing the rates of $\KL$ and
$\KS$ decay into $\pi^+\pi^-$ and $\pi^0\pi^0$. This quantity has been measured
precisely~\cite{PDG2016,Batley:2002gn,Abouzaid:2010ny} but is
difficult to predict theoretically, as it receives dominant
contributions from two four-quark operators (denoted $Q_6$ and
$Q_8$ in the framework of the effective Hamiltonian) that largely
cancel each other. A lattice QCD evaluation of all the bag
parameters needed has recently~been performed \cite{Bai:2015nea},
suggesting a discrepancy between 2$\sigma$ and 3$\sigma$ with
respect to SM
expectations~\cite{Bai:2015nea,Buras:2015jaq,Buras:2015xba,Kitahara:2016nld}.
This interesting but challenging issue definitely calls for
estimations of the relevant bag parameters from other lattice QCD
collaborations.

\boldsubsection{Lepton Flavour Universality}\label{Sec:LFU}

The metrology of the CKM parameters discussed above relies on
modes that can be predicted accurately in the SM and provide
information about its parameters. However, it mixes modes with
different sensitivities to physics beyond the SM: on one hand,
flavour-changing charged currents, such as semileptonic decays,
which are dominated by tree-level processes in the SM, and on the other
hand, flavour-changing neutral currents, such as neutral-meson
mixing, which are mediated by loop processes in the SM.
Additional, rare processes that are not expected to provide
further constraints on the parameters of the SM can probe some of
the underlying hypotheses at the core of this theory. More details
can be found in a previous volume of this
journal~\cite{Blake:2015tda}.

A particularly topical example is lepton flavour universality. In
both flavour-changing charged and neural currents, the weak
interaction at play deals with lepton flavours in a universal
manner, whereas quarks are treated on a different footing due to
the CKM matrix. This universality of lepton couplings is assumed
when determining the CKM parameters, in particular to combine
results from semileptonic and leptonic decays that involve
$e,~\mu$, and/or $\tau$ leptons.

Recently, LHCb and the \B factories found interesting
hints of violation of lepton flavour universality in both
flavour-changing charged and neural
currents~\cite{LHCb-PAPER-2014-024}. The measurements in charged
currents between \decay{\B}{D^{(*)}\tau\nu} and
\decay{\B}{D^{(*)}\ell\nu}, where
$\ell=\mu,e$~\cite{Lees:2012xj,Lees:2013uzd,Huschle:2015rga,Abdesselam:2016cgx,LHCb-PAPER-2015-025,Hirose:2016wfn},
indicate that the ratios $R(D)$ and $R(D^*)$ exceed SM predictions
by 1.9$\sigma$ and 3.3$\sigma$, respectively, leading to a combined
discrepancy with the SM at 4.0$\sigma$~\cite{HFAG}:
\begin{equation}
R_{D(^*)}=\frac{\mathrm{Br}(B\to D^{(*)}\tau\nu)}{\mathrm{Br}(B\to
D^{(*)}\ell\bar\nu_\ell)}\,.
\end{equation}
The individual branching ratios are consistent with a 15\%
enhancement for $b\to c\tau\bar\nu_\tau$ compared with SM
expectations. Several similar measurements, notably from LHCb, are
ongoing and should provide a clearer picture in the near future.

The violation of lepton flavour universality has also been
investigated for the flavour-changing neutral-current (FCNC)
transition $b\to s\ell^+\ell^-$ at several experiments. LHCb~\cite{LHCb-PAPER-2014-024} measured the
observable $R_K=\mathrm{Br}(B\to K\mumu)/\mathrm{Br}(B\to Ke^+e^-)$ in the dilepton mass range from 1
to 6 GeV$^2$ as $0.745^{+0.090}_{-0.074}\pm 0.036$, corresponding to a $2.6\sigma$ tension with its SM value, which is predicted to be equal
to one (to high accuracy). This violation has also been studied in $B\to K^*\ell^+\ell^-$ transitions, with $R_{K^{*0}}$ measured in two low-$q^2$ bins with deviations from the SM between 2.2 and 2.5$\sigma$~\cite{LHCb-PAPER-2017-013}.
Other recent experimental results have
shown interesting deviations from the SM in the muon sector. The
LHCb analysis~\cite{LHCb-PAPER-2015-051} of the decay
$\decay{\Bd}{\Kstarz\mumu}$ reports an $\sim3\sigma$ anomaly in
two large $K^*$ recoil bins of the angular observable
$P_5^\prime$~\cite{DescotesGenon:2012zf}. This report was subsequently
confirmed by the Belle experiment~\cite{Abdesselam:2016llu} with
the hint that it would arise in \decay{\bquark}{\squark\mumu} but
not in \decay{\bquark}{\squark \epem}~\cite{Wehle:2016yoi}.
Finally, the LHCb results for the branching ratio of several
\decay{\bquark}{\squark\mumu} decays exhibit deviations at low
dilepton
masses~\cite{LHCb-PAPER-2013-017,LHCb-PAPER-2014-006,LHCb-PAPER-2015-009,LHCb-PAPER-2015-023}.

Confirmation of these deviations from lepton flavour universality would be an unambiguous sign of physics beyond the SM. It would
also have consequences for the constraints described above, especially
those in Section~\ref{Sec:Moduli}, which are determined using
leptonic and semileptonic decays. Most analyses assume lepton
universality, a hypothesis that would need to be revisited (see
Section~\ref{Sec:globalLFUV} for more detail).

\section{GLOBAL ANALYSES}\label{Sec:global}

\boldsubsection{Determination of CKM Parameters}
The following subsections describe how the above-mentioned individual constraints can be combined to constrain the CKM parameters.
\boldsubsubsection{Statistical approaches to global analyses}

The individual constraints presented above must be combined in
order to obtain statistically meaningful constraints on the CKM
parameters. The problem can be described as a series of
observables (e.g., branching ratios of leptonic and semileptonic
decays, mass difference for neutral mesons) depending on
theoretical parameters. Some of these are of interest
($A,\lambda,\bar\rho,\bar\eta$); the others are called
nuisance parameters (e.g., decay constants, form factors,
quark masses). The
primary goal of statistical analysis is to determine the confidence
intervals for the CKM parameters (and other fundamental parameters
for models beyond the SM). The accuracy of the determination of
the CKM parameters thus depends on the precision of the
experimental measurements and on the theoretical computations of
the nuisance parameters. Currently, global analyses are limited
mainly by the latter, which are obtained mostly from QCD lattice simulations that
consider a discretised version of QCD on a finite grid and compute
correlators through Monte Carlo integrations over gluon gauge
configurations. Due to the remarkable improvement in computing
power and algorithms over recent decades, these computations
are now dominated mainly by systematic uncertainties
(extrapolation in lattice spacing, volume and quark masses,
renormalisation).

Therefore, a global analysis requires both a general statistical
framework and a specific model for systematic uncertainties.
Frequentist and Bayesian approaches have been proposed to deal
with such analyses: The former defines probability as the outcome
of repeated trials/measurements in the limit where their number
becomes infinite, and the latter considers them as a subjective
degree of credibility given by the observer to each possible
result. The choice between the two approaches is the subject of
considerable discussion in the literature (a specific discussion
regarding the CKM case can be found in
References~\cite{Battaglia:2003in,Charles:2006vd,Bona:2007qta,Charles:2007yy}).
The frequentist approach has been adopted by the CKMfitter
Group~\cite{Hocker:2001xe,Charles:2004jd}, whereas the Bayesian
approach is used by the UTfit Group~\cite{Ciuchini:2000de}.

Another issue, the models for systematic uncertainties, is also a
matter of debate. For lack of a better choice, and even though they
are not of a statistical nature by definition, systematic uncertainties are often
described with the same model as statistical uncertainties, for
instance, in the case of the UTfit group~\cite{Ciuchini:2000de}.
Alternative treatments consist of determining sets of confidence
intervals for specific values of the systematic uncertainties
before combining them in unified confidence intervals [the scan
method~\cite{Eigen:2013cv}] or building dedicated models for
likelihoods and \textit{p} values treating a range of values for the
systematic uncertainties on an equal footing [the Rfit model used by
the CKMfitter Collaboration~\cite{Charles:2004jd}]. This choice
has an effect not only when performing the global fit itself but
also when choosing inputs by averaging measurements or
computations from different groups. A more detailed discussion of
the various models for theoretical uncertainties can be found in
Reference~\cite{Charles:2016qtt}.

\boldsubsubsection{Determination of the CKM parameters and consistency tests}

\begin{figure}\centering
\includegraphics[width=\textwidth]{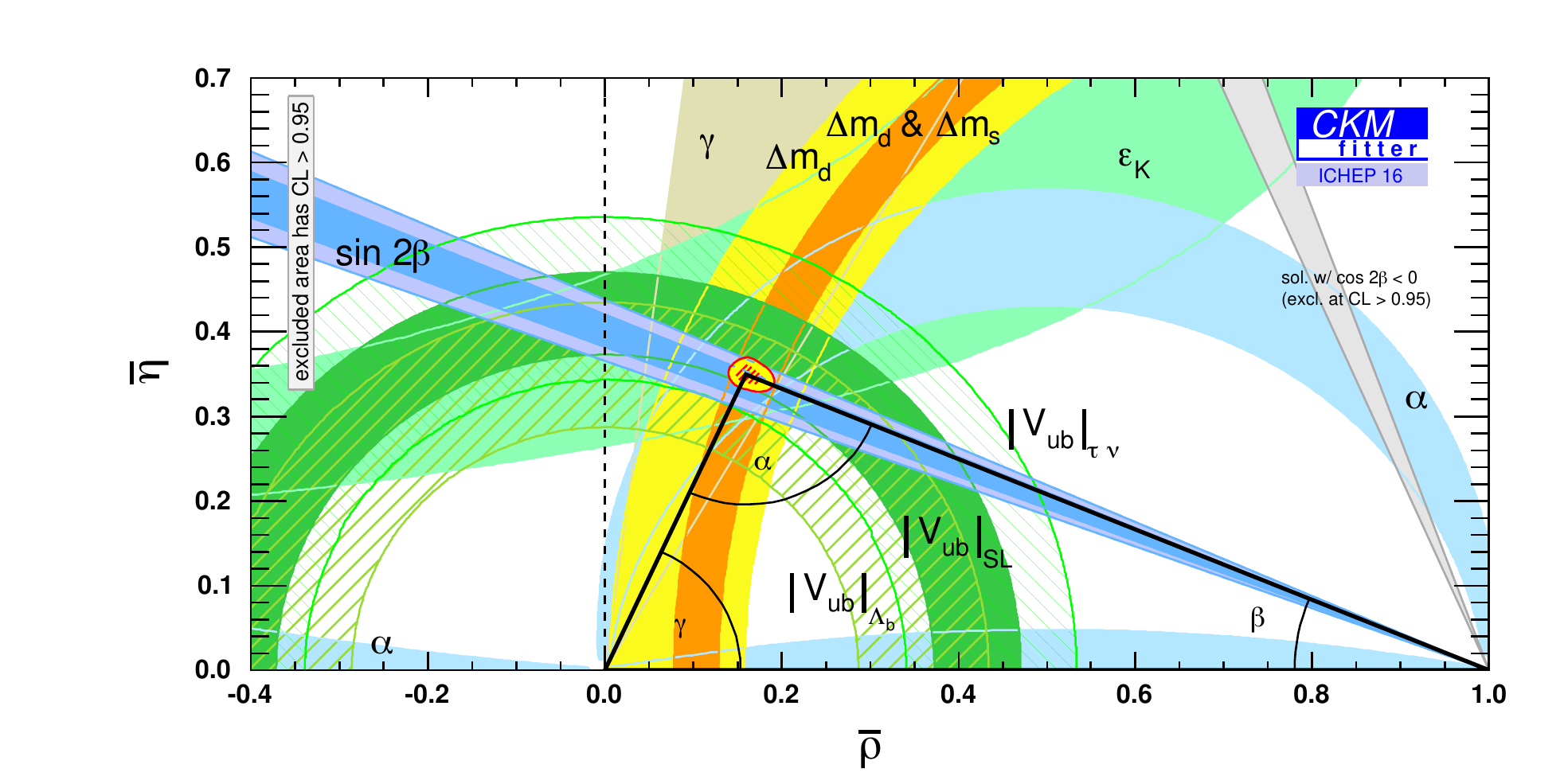}
\caption{Status of the CKM unitarity triangle fit in
$(\bar\rho,\bar\eta)$ in summer 2016. Regions outside the coloured
areas have ${\rm CL} > 95.45\%$. For the combined fit, the yellow
area inscribed by the contour line represents points with ${\rm
CL} < 95.45\%$. The shaded area inside this region represents
points with ${\rm CL} < 68.3\%$~\cite{Charles:2004jd}.}\label{Fig:CKMFit1}
\end{figure}

For illustrative purposes, we use the results obtained by the CKMfitter
Group, based on the results available at the time of the 2016 International Conference on High Energy Physics (ICHEP)~\cite{Charles:2004jd}. 
Figure~\ref{Fig:CKMFit1} depicts the current situation regarding the
global fit in the $(\bar\rho,\bar\eta)$ plane. Table~\ref{tab:expinputs} lists the input parameters. As indicated in Section~\ref{Sec:UT},
this result could be cast into other UTs.

\begin{sidewaystable}
\caption{Constraints used for the global fit and the main inputs
involved. When two uncertainties are quoted, the first one is
statistical and the second is systematic. The lattice inputs and the averaging method used are
discussed by the CKMfitter Group~\cite{Charles:2004jd}, along with additional
theoretical inputs (quark masses, strong coupling constant,
short-distance QCD corrections for meson mixing). For a review of
lattice inputs, see Reference~\cite{Aoki:2016frl}. Abbreviations:\ ADS, Atwood--Dunietz--Soni; GGSZ, Giri--Grossman--Soffer--Zupan; GLW, Gronau--London--Wyler; OPE, operator product expansion.\label{tab:expinputs}}
\resizebox{\textheight}{!}{\begin{tabular}{lllllllll}
CKM & Process & \multicolumn{4}{c|}{Observables} & \multicolumn{3}{c}{Theoretical inputs}\\
\hline
$|V_{ud}|$ & $0^+\to 0^+$ transitions & $|V_{ud}|_{\rm nucl}$&=& $0.97425\pm 0\pm 0.00022$ & \cite{Hardy:2014qxa} & \multicolumn{3}{c}{Nuclear matrix elements} \\
\hline
$|V_{us}|$ & $K\to\pi\ell\nu$ & $|V_{us}|_{\rm SL}f_+^{K\to\pi}(0)$&=& $ 0.2165\pm0.0004 $ & \cite{PDG2016} & $f_+^{K\to\pi}(0)$&=& $0.9681\pm 0.0014\pm 0.0022$\\
& $K\to e\nu_e$ & ${\cal B}(K\to e\nu_e)$&=&$(1.581\pm0.008)\times 10^{-5}$ & \cite{PDG2016} & $f_K$&=& $155.2\pm0.2\pm0.6 $ MeV \\
& $K\to \mu\nu_\mu$ & ${\cal B}(K\to \mu\nu_\mu)$&=& $0.6355 \pm 0.0011$ & \cite{PDG2016}\\
& $\tau \to K \nu_\tau$ & ${\cal B}(\tau \to K\nu_\tau)$&=&$(0.6955 \pm 0.0096)\times 10^{-2}$ & \cite{PDG2016}\\
\hline
$\frac{|V_{us}|}{|V_{ud}|}$ & $K\to \mu\nu/\pi\to\mu\nu$ & $\displaystyle \frac{{\cal B}(K\to \mu\nu_\mu)}{{\cal B}(\pi \to \mu\nu_\mu)}$ &=&$1.3365 \pm 0.0032$ & \cite{PDG2016} & $f_K/f_\pi$&=&$1.1959 \pm 0.0010\pm0.0029$\\
& $\tau\to K\nu/\tau \to \pi\nu$ & $\displaystyle \frac{{\cal B}(\tau \to K\nu_\tau)}{{\cal B}(\tau \to \pi\nu_\tau)}$ &=& $(6.43 \pm 0.09)\times 10^{-2}$ & \cite{PDG2016} \\
\hline
$|V_{cd}|$ & $\nu N$ & $|V_{cd}|_{\nu N}$ &=& $0.230\pm 0.011$ & \cite{PDG2016}\\
& $D\to \mu\nu $ & ${\cal B}(D\to \mu\nu)$ &=& $(3.74\pm0.17) \times 10^{-4}$ & \cite{HFAG} & $f_{D_s}/f_D$&=&$1.175 \pm 0.001\pm0.004$\\
& $D\to \pi\ell\nu $ & $|V_{cd}|f_+^{D\to \pi}(0)$ &=& $0.1425 \pm 0.0019$ & \cite{Besson:2009uv,Widhalm:2006wz} & $f_+^{D\to \pi}(0)$&=&$0.666\pm 0.020\pm 0.048 $\\
\hline
$|V_{cs}|$ & $W\to c\bar{s}$ & $|V_{cs}|_{W\to c\bar{s}}$ &=& $0.94^{+0.32}_{-0.26}\pm 0.13$ & \cite{PDG2016}\\
& $D_s\to \tau\nu$ & ${\cal B}(D_s\to \tau\nu)$&=& $(5.55\pm0.24) \times 10^{-2}$ & \cite{HFAG} & $f_{D_s}$ &=& $248.2\pm 0.3 \pm 1.9$ MeV\\
& $D_s\to \mu\nu$ & ${\cal B}(D_s\to \mu\nu_\mu)$&=& $(5.57\pm0.24)\times 10^{-3}$ & \cite{HFAG}\\
& $D\to K\ell\nu $ & $|V_{cs}|f_+^{D\to K}(0)$ &=& $0.7226\pm 0.0034$ & \cite{Aubert:2007wg,Besson:2009uv,Widhalm:2006wz} & $f_+^{D\to K}(0)$&=&$ 0.747\pm0.011\pm0.034$\\
\hline
$|V_{ub}|$ & Semileptonic $B$ & $|V_{ub}|_{\rm SL}$ &=& $(3.98 \pm 0.08 \pm 0.22)\times 10^{-3}$ & \cite{HFAG,Charles:2004jd} & \multicolumn{3}{c}{Form factors, shape functions}\\
& $B\to \tau\nu$ & ${\cal B}(B\to\tau\nu)$ &=& $(1.08\pm0.21) \times 10^{-4}$ & \cite{HFAG} & $f_{B_s}/f_B$&=& $1.205\pm 0.003 \pm 0.006 $\\
\hline $|V_{cb}|$ & Semileptonic $B$
 & $|V_{cb}|_{\rm SL}$ &=& $(41.00\pm 0.33 \pm 0.74 )\times 10^{-3}$ & \cite{HFAG}
 & \multicolumn{3}{c}{Form factors, OPE matrix elements}\\
$|V_{ub}/V_{cb}|$ & Semileptonic $\Lb$
 & $\frac{\BR(\Lb\to p\mu^-\bar\nu_\mu)_{q^2>15}}{\BR(\Lb\to \Lc\mu^-\bar\nu_\mu)_{q^2>7}}$ &=& $(0.944\pm 0.081) \times 10^{-2}$ & \cite{LHCb-PAPER-2015-013}
 & \multicolumn{3}{c}{$\frac{\zeta(\Lb\to p\mu^-\bar\nu_\mu)_{q^2>15}}{\zeta(\Lb\to \Lc\mu^-\bar\nu_\mu)_{q^2>7}}= 1.471\pm 0.096\pm 0.290$} \\
\hline $\alpha$ & $B\to\pi\pi$, $\rho\pi$, $\rho\rho$
 & \multicolumn{3}{c}{Branching ratios, $\CP$ asymmetries} & \cite{HFAG}
 & \multicolumn{3}{c}{Isospin symmetry}\\
 \hline
$\beta$ & $B\to (c\bar{c}) K$
 & $\sin(2\beta)_{[c\bar{c}]}$ &=& $0.691 \pm 0.017$
 & \cite{HFAG} & \multicolumn{3}{c}{Penguin neglected}\\
\hline $\gamma$ & $B\to D^{(*)} K^{(*)}$
 & \multicolumn{3}{c}{Inputs for the three methods}
 & \cite{HFAG}& \multicolumn{3}{c}{GGSZ, GLW, ADS methods} \\
 \hline
$\phi_s$ & $B_s\to J/\psi (KK, \pi\pi)$ & $\phi_s$ &=& $-0.030\pm
0.033$
 & \cite{HFAG}& \multicolumn{3}{c}{Penguin neglected}
 \\
 \hline
$V_{tq}^*V_{tq'}$ & $\Delta m_d$
 & $\Delta m_d$ &=& $0.5065 \pm 0.0019$ ps${}^{-1}$
 & \cite{HFAG}
 & $\hat{B}_{B_s}/\hat{B}_{B_d}$ &=& $1.007 \pm 0.014\pm0.014$\\
 & $\Delta m_s$ & $\Delta m_s$ &=& $17.757\pm0.021$ ps${}^{-1}$
 & \cite{HFAG}
 & $\hat{B}_{B_s}$&=& $1.320\pm0.016\pm0.030$\\
 & $B_s\to \mu\mu$ & ${\cal B}(B_s\to\mu\mu)$ &=& $(2.8^{+0.7}_{-0.6})\times 10^{-9}$
 & \cite{LHCb-PAPER-2014-049} & $f_{B_s}$ &=& $225.1\pm 1.5\pm 2.0$ MeV \\
 \hline
$V_{td}^*V_{ts}$
 & $\epsilon_K$ & $|\epsilon_K|$ &=& $(2.228\pm0.011)\times 10^{-3}$
 & \cite{PDG2016}
 &$\hat{B}_K$&=& $0.7567\pm0.0021\pm0.0123$\\
$V_{cd}^*V_{cs}$ & &&&&& $\kappa_\epsilon$&=&
$0.940\pm0.013\pm0.023 $\\
\hline
\end{tabular}}
\end{sidewaystable}

Some comments are in order before we discuss the metrology of the
parameters. There exists a unique preferred region defined by the
entire set of observables under consideration in the global fit.
In Figure 6, this region is represented by the yellow surface inscribed by the
red contour line for which the values of $\bar\rho$ and $\bar
\eta$ with a $p$ value such that $1-p < 95.45\%$. The goodness of
the fit must be assessed in relation to the model used to
describe the theoretical uncertainties. If all of the inputs'
uncertainties are assumed to be statistical in nature, and if they can be
combined in quadrature, the corresponding minimal $\chi^2$ has a
$p$ value of $20\%$ (i.e., 1.3$\sigma$). The following values for
the four parameters describing the CKM matrix are obtained:

\begin{equation} \label{eq:CKMparams2016}
A= 0.825\aerr{+0.007}{-0.012}, \quad
\lambda=0.2251\aerr{+0.0003}{-0.0003},\quad \bar\rho =
0.160\aerr{+0.008}{-0.007}, \quad \bar\eta =
0.350\aerr{+0.006}{-0.006}.
\end{equation}
The overall consistency is striking when comparing constraints
from tree-mediated (leptonic and semileptonic decays) and loop-mediated
(e.g., neutral-meson mixing) processes, as well as processes
requiring \CP violation (such as nonvanishing \CP asymmetries)
with respect to processes taking place even if \CP were conserved
(such as leptonic and semileptonic decays) (Figure~\ref{Fig:CKMFit1}). The consistency observed among the constraints allows one to perform the metrology of
the CKM parameters and to give predictions for any CKM-related
observable within the SM. Each comparison between the prediction
issued from the fit and the corresponding measurement constitutes
a null test of the SM hypothesis.

\begin{figure}\centering
\includegraphics[width=0.8\textwidth]{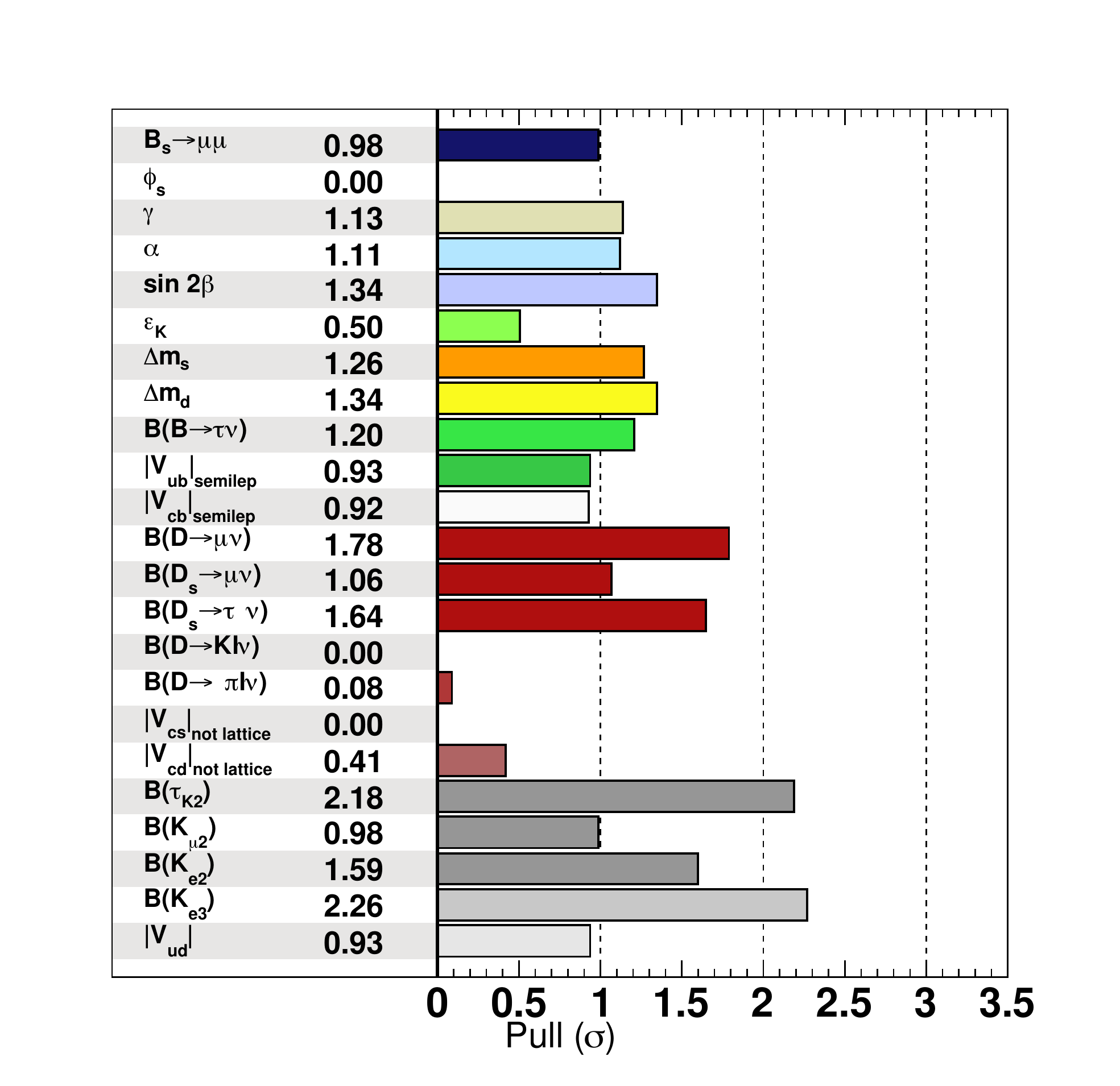}
\caption{Pulls for the global fit in summer 2016, as defined in
Reference~\cite{Lenz:2010gu} and by the CKMfitter Group~\cite{Charles:2004jd}. Each pull amounts to the absolute
difference between the predicted and measured values, divided by the
experimental uncertainty when the latter is large compared with the
uncertainty of the prediction~\cite{CDF/ANAL/PUBLIC/5776}.}\label{Fig:CKMFit2}
\end{figure}

Figure~\ref{Fig:CKMFit2} shows some of the corresponding pulls, demonstrating that there is no sign of
discrepancy with this set of inputs. In particular, recent
discrepancies related to $B(B\to\tau\nu)$, $\sin(2\beta)$,
$\varphi_s$ \cite{Lenz:2010gu}, $\Vcb$,
$\epsilon_K$~\cite{Lunghi:2008aa,Buras:2008nn}, or $\Delta
m_{d,s}$~\cite{Bazavov:2016nty} do not appear, either because
recent changes in the experimental inputs or because of the dependence of
these discrepancies on the statistical treatment and the modelling
of systematic uncertainties.

Unitarity tests using direct determination of individual matrix
elements (without resorting to unitarity) can also be performed by
checking that the sum of their squares equals unity. For the
first two rows of the CKM matrix, the following results are
obtained
\begin{eqnarray}
|\Vud|^2_{\rm meas}+|\Vus|^2_{\rm meas}+|\Vub|^2_{\rm meas}-1&=&-0.0006\aerr{+0.0006}{-0.0002},\\
|\Vcd|^2_{\rm meas}+|\Vcs|^2_{\rm meas}+|\Vcb|^2_{\rm
meas}-1&=&-0.0034\aerr{+0.0048}{-0.0026},
\end{eqnarray}
where each ``measured'' value includes all semileptonic and
leptonic direct determinations of a given CKM matrix element (an
average of inclusive and exclusive semileptonic measurements is
used for the semileptonic input for $|\Vub|$ and $|\Vcb|$). No
deviation from unitarity is observed. There is no direct
determination of $|\Vtd|$ and $|\Vts|$ (they are obtained from
$\Delta F=2$ loop processes), and there is no accurate direct
determination of $|\Vtb|$~\cite{PDG2016}; thus, no equivalent test can be performed for the third row or any of the columns of the
CKM matrix. Similarly, the value of $\alpha+\beta+\gamma$ cannot
be probed directly, because the determination of $\alpha$ from $B\to
\pi\pi,\pi\rho,\rho\rho$ already assume unitarity.

\begin{table}\centering
\caption{A few predictions from the global fit (indirect, i.e.,
not including direct determinations of these quantities) compared
with direct determinations. The top panel corresponds to
experimental inputs; the bottom panel to inputs from lattice QCD
computations. In the case of $\mathrm{Br}(\Bs\to\mumu)$, the value corresponds
to the value before integration over time, i.e., removing the
effect of $\Delta \Gamma_s$. For ${\rm Br}(B^0\to\mu^+\mu^-)$, an upper bound is available, but the statistical significance is too low to quote a measurement in the right-hand column~\cite{Charles:2004jd}.}
\begin{tabular}{lll}
\hline
Quantity & Fit prediction & Direct determination\\
\hline
$\alpha\ (^\circ)$ & $92.1\aerr{+1.5}{-1.1}$ & $88.8\aerr{+2.3}{-2.3}$ \\
$\beta\ (^\circ)$ & $23.7\aerr{+1.1}{-1.0}$ & $21.8\aerr{+0.7}{-0.7}$\\
$\gamma\ (^\circ)$ & $65.3\aerr{+1.0}{-2.5}$ & $72.1\aerr{+5.4}{-5.8}$\\
$\varphi_s\ (\rm rad)$ & $-0.0370\aerr{+0.0006}{-0.0007}$ & $-0.030\pm 0.033$ \\
$\mathrm{Br}(\Bs\to\mumu)\times 10^9$ & $3.36\aerr{+0.07}{-0.19}$ & $2.62\aerr{+0.66}{-0.56}$ \\
$\mathrm{Br}(\Bd\to\mumu)\times 10^{11}$ & $9.55\aerr{+0.25}{-0.44}$ & $-$ \\
$\Vub \times 10^3$ & $3.60\aerr{+0.10}{-0.10}$ & $3.98\pm 0.08\pm 0.22$\\
$\Vcb \times 10^3$ & $42.2\aerr{+0.7}{-0.7}$ & $41.00\pm 0.33 \pm 0.74$ \\
\hline
$f_K$ & $0.15652\aerr{+0.00013}{-0.00020}$ & $0.1552\pm 0.0002\pm 0.0006$ \\
$f_K/f_\pi$ & $1.1965\aerr{+0.0021}{-0.0063}$ & $1.1959 \pm 0.0010\pm0.0029$ \\
$f_+^{K\to \pi}(0) $& $0.9602\aerr{+0.0020}{-0.0025}$ & $0.9681\pm 0.0014\pm 0.0022$ \\
$\hat{B}_K$ & $0.79\aerr{+0.17}{-0.11}$ & $0.7567\pm0.0021\pm0.0123$\\
$f_{D_s}$ (GeV) & $0.2512\aerr{+0.0032}{-0.0032}$ & $0.2482\pm 0.0003 \pm 0.0019$ \\
$f_{D_s}/f_D$ & $1.226\aerr{+0.029}{-0.027}$ & $1.175 \pm 0.001\pm0.004$ \\
$f_+^{D\to \pi}(0)$ &$0.633\aerr{+0.009}{-0.008}$ & $0.666\pm 0.020\pm 0.048$ \\
$f_+^{D\to K}(0)$ & $0.742\aerr{+0.004}{-0.004}$ & $0.747 \pm 0.011\pm0.034$\\
$f_{\Bs}$ (GeV) & $0.226\aerr{+0.004}{-0.005}$ & $0.2251\pm 0.0015\pm 0.0020$ \\
$f_{\Bs}/f_B$ & $1.243\aerr{+0.027}{-0.020}$ & $1.205\pm 0.003 \pm 0.006$\\
$\hat{B}_{\Bs}$ & $1.332\aerr{+0.040}{-0.067}$ & $1.320\pm0.016\pm0.030$ \\
$\hat{B}_{\Bs}/\hat{B}_{\Bd}$ & $1.114\aerr{+0.046}{-0.047}$ & $1.007 \pm 0.014\pm0.014$\\
\hline
\end{tabular}
\label{Tab:IndirectPredictions}
\end{table}

The global fit also provides indirect predictions (i.e., not
including direct measurements of these quantities) for quantities
of interest, either measured experimentally or determined from
lattice QCD simulations (Table~\ref{Tab:IndirectPredictions}).
A similar level of accuracy is achieved for some observables in
both their direct determinations and their indirect prediction.
Improving their measurement will have only a limited impact on the
fit, unless the central value differs significantly from the
global fit expectations (which would then require a fine
understanding of all sources of uncertainties of the
measurements). Other quantities are still far from being measured
as accurately as their prediction from the global fit. Their
measurements can help further constrain the CKM parameters, and
they still leave room for unexpected deviations from the SM
picture emerging from the global fit.

\boldsubsection{Analyses of Deviations from the CKM Paradigm}\label{Sec:globalNP}

Quark flavour physics provides both stringent tests of the SM and
significant constraints on NP models. However, the above-described processes
used to determine the CKM parameters show good overall
consistency within the SM, and thus lead to upper bounds on
additional NP contributions. Additional processes suffering from
larger theoretical or experimental uncertainties must therefore be
included in the global analyses in order to probe physics beyond
the SM.

Although specific NP models could be directly compared with
experimental results, it is natural to
consider effective approaches for flavour processes. The short-distance dynamics is
encoded in Wilson coefficients multiplied by operators describing
the transition on long distances~\cite{Buchalla:1995vs}, given that
these flavour processes take place at significantly lower energies
than the NP degrees of freedom of interest. NP affects
the values of the Wilson coefficients. The structure of the
operators affected (e.g., vector, scalar) provides a hint of the
type of NP at play, and the deviations of the Wilson coefficients
from SM expectations provide an idea of the energy scales and
coupling constants involved. In any case (specific NP models or
general effective approaches), the above constraints must be
reconsidered in order to learn whether they can be used to
determine CKM parameters, constrain NP contributions, or neither.

We discuss two topical examples in the following subsections. The first is NP arising in
$\Delta F=2$ processes, and the second is NP violating lepton flavour
universality in $\Delta F=1$ processes.
\boldsubsubsection{New Physics in $\Delta F = 2$}\label{Sec:DeltaF2NP}

As discussed elsewhere~\cite{Soares:1992xi,Goto:1995hj,Silva:1996ih,Grossman:1997dd,Bona:2005eu,Ligeti:2006pm,Bona:2007vi,Lenz:2010gu,Lenz:2012az,Charles:2013aka}
and in Section~\ref{Sec:DeltaF2BdBs}, neutral-meson mixing is a
particularly interesting probe of NP. The evolution of the
$B_q\bar{B}_q$ system is described through a quantum-mechanical
Hamiltonian $H=M^q-i\Gamma^q/2$ as the sum of two Hermitian
``mass'' and ``decay'' matrices, so that
$B^0_{(s)}$--$\Bbar^0_{(s)}$ oscillations involve the off-diagonal
elements $M_{12}^q$ and $\Gamma_{12}^q$, respectively. The three
physical quantities $|M_{12}^q|$, $|\Gamma_{12}^q|$, and
$\varphi_q=\arg(-M_{12}^q/\Gamma_{12}^q)$ can be determined from
the mass difference $\dm_q\simeq 2|M_{12}^q| $ among the
eigenstates, their width difference $\Delta\Gamma_q \simeq 2\,
|\Gamma_{12}^q| \cos \varphi_q$, and the semileptonic $\CP$
asymmetry $a^q_{\rm SL} = {\rm Im} {\Gamma_{12}^q}/{M_{12}^q}=
{\Delta\Gamma_q}/{\dm_q} \tan \varphi_q$. Resulting from box
diagrams with heavy (virtual) particles, $M_{12}^q$ is expected to
be especially sensitive to NP~\cite{Lenz:2010gu}, so that the two
complex parameters $\Delta_s$ and $\Delta_d$, defined as
\begin{equation}
M_{12}^q \!\equiv\! M_{12}^{\text{SM},q} \Delta_q, \quad \Delta_q
\equiv |\Delta_q| e^{i \varphi^\Delta_q}, \quad q=d,s,
\label{defdel}
\end{equation}
can differ substantially from the SM value $\Delta_s=\Delta_d=1$.

Importantly, the NP phases $\varphi^\Delta_{d,s}$ not only
affect $a^{d,s}_{\rm SL}$ but also shift the $\CP$ phases
extracted from the mixing-induced $\CP$ asymmetries in
\decay{\Bz}{\jpsi\KS} and \decay{\Bs}{\jpsi\phi} to
$2\beta+\varphi^\Delta_d$ and $2\beta_s-\varphi^\Delta_s$,
respectively. If it is assumed that NP enters only through the two
parameters $\Delta_d$ and $\Delta_s$, the CKM paradigm is still
valid to analyse $\Delta F=1$ quark flavour transitions. By contrast, the $\Delta F=2$ transitions previously used to
determine the CKM parameters must be reinterpreted as constraints
on $\Delta_d$ and $\Delta_s$ [namely $\Delta m_d$, $\Delta m_s$,
$\sin(2\beta)$ and $\alpha$].

There has been a great deal of interest in such NP scenarios triggered by
deviations observed first in early measurements from CDF and
\dzero on the \Bs mixing angle $\varphi_s$, and later after D0
quoted values of the like-sign dimuon asymmetry $a_{\rm SL}$
(measuring a linear combination of $a^d_{\rm SL}$ and $a^s_{\rm
SL}$). However, as discussed in Section~\ref{Sec:DeltaF2BdBs},
later measurements of the individual semileptonic \CP asymmetries
and mixing angles for \Bd and \Bs mesons have not been able to
explain the D0 measurement, as they showed good agreement with
SM expectations.

\begin{figure}\centering
\includegraphics[width=0.49\textwidth]{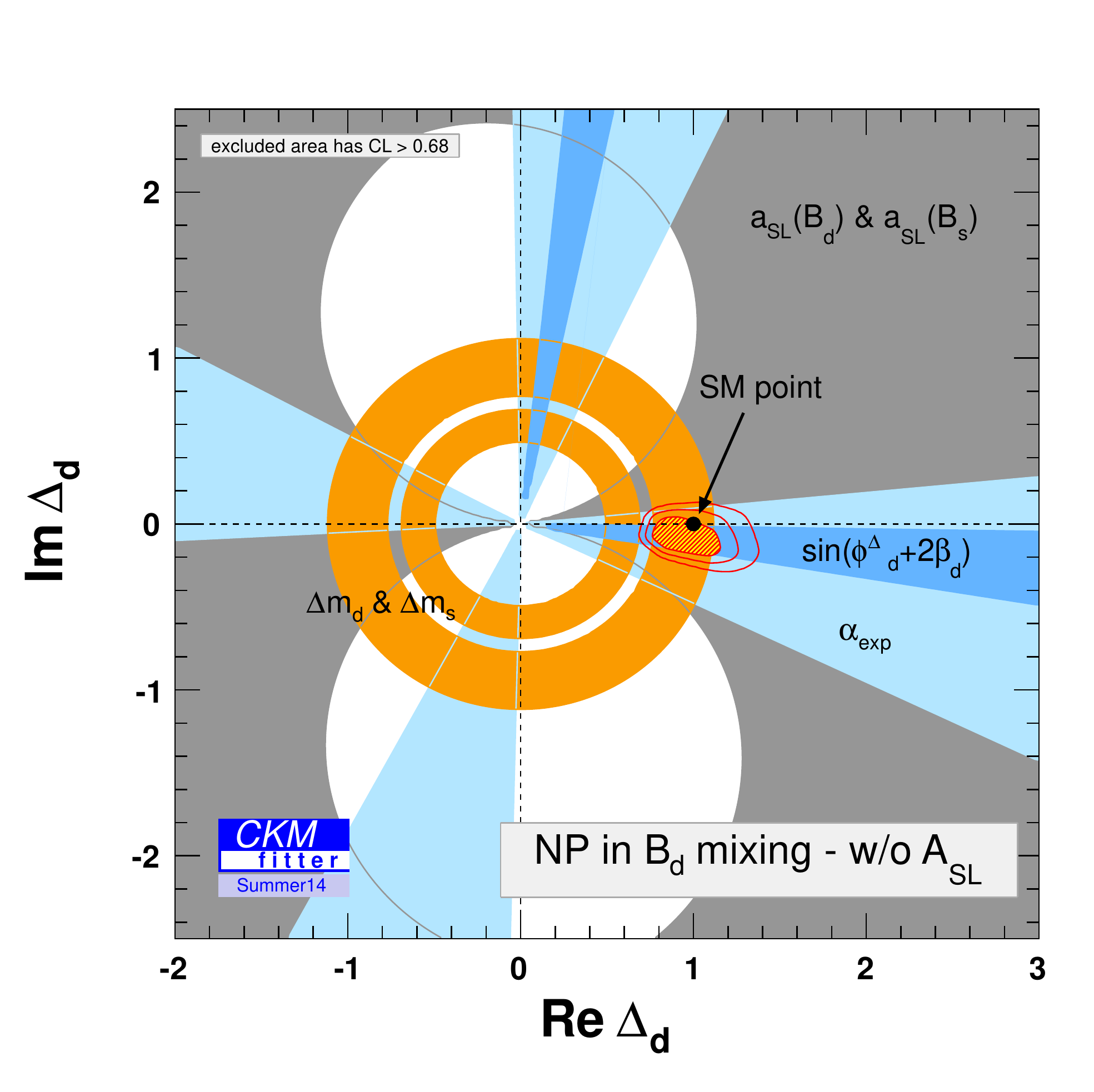}\hskip 0.01\textwidth
\includegraphics[width=0.49\textwidth]{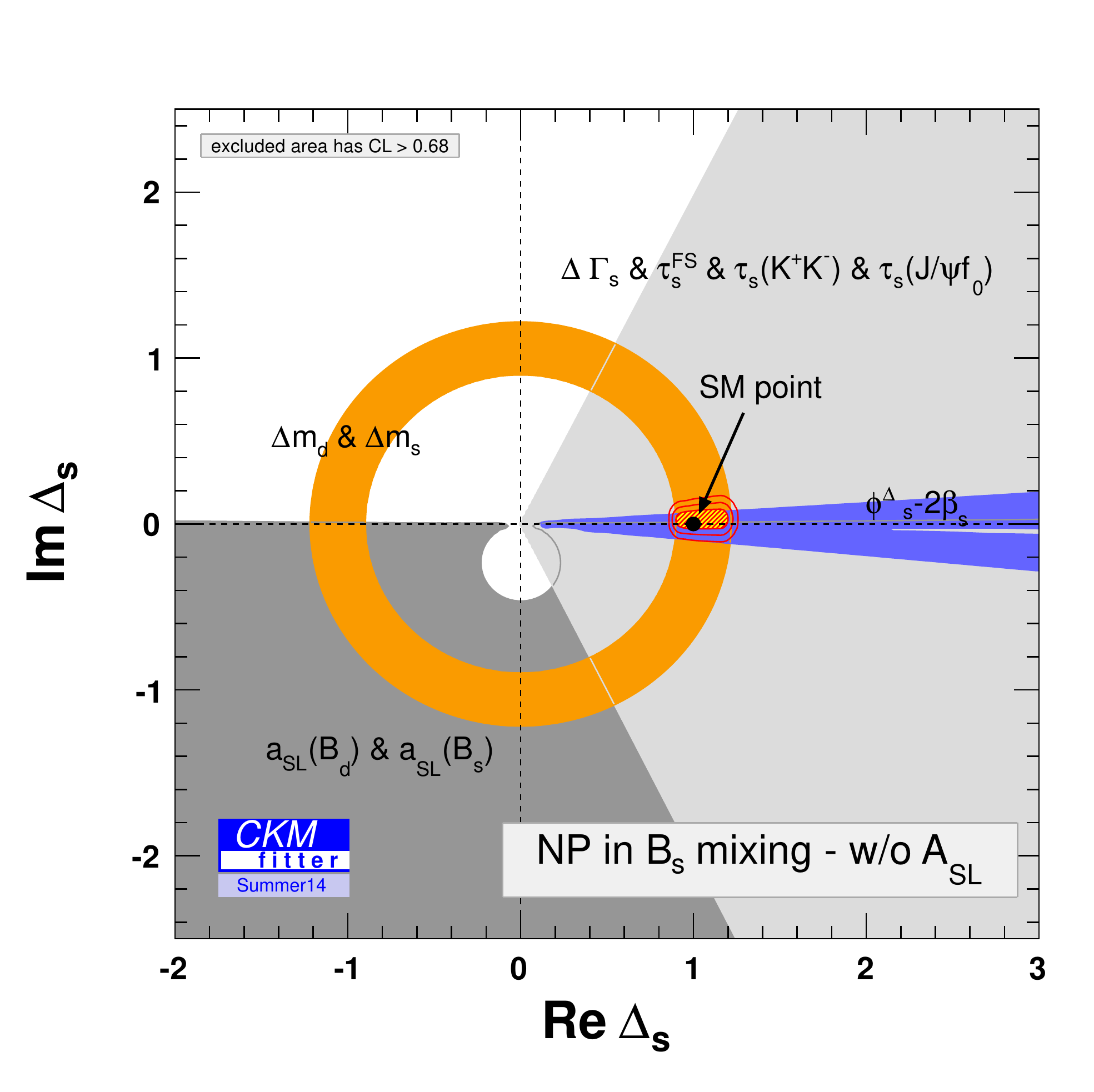}
\caption{Complex parameters (\textit{a}) $\Delta_d$ and (\textit{b})
$\Delta_s$ describing New Physics (NP) in $\Delta F=2$ (Scenario I, not
including $a_{\rm SL}$). The coloured areas represent regions with
$1-p < 68.3\%$ for the individual constraints. The red area represents
the region with $1-p < 68.3~\%$ for the combined fit, with the two
additional contours delimiting the regions with $1-p < 95.45~\%$
and $1-p < 99.73~\%$. Abbreviation:\ SM, Standard Model~\cite{Charles:2015gya}.}\label{Fig:DeltaF2}
\end{figure}

Simultaneous fits of the CKM parameters and the NP parameters
$\Delta_d$, and $\Delta_s$ have been
performed~\cite{Lenz:2010gu,Lenz:2012az} in different generic
scenarios in which NP is confined to $\Delta F=2$ flavour-changing
neutral currents. The most recent update~\cite{Charles:2015gya}
used data up to summer 2014. The two complex NP parameters
$\Delta_d$ and $\Delta_s$ are not sufficient to absorb the
discrepancy between the D0 measurement of $a_{\rm SL}$ and the
rest of the global fit~\cite{Charles:2015gya}. Without $a_{\rm
SL}$, the fit including NP in $\Delta F=2$ is good, but the
improvement with respect to the SM is limited. In the case of the
so-called scenario I ($\Delta_s$ and $\Delta_d$ independent), the
following values are obtained:

\begin{equation}
\Delta_d=(0.94\aerr{+0.18}{-0.15})+i(-0.12\aerr{+0.12}{-0.05}) \qquad
\Delta_s=(1.05\aerr{+0.14}{-0.13})+i(0.03\aerr{+0.04}{-0.04}),
\end{equation}
together with the following values of the CKM parameters:

\begin{equation}
A= 0.790\aerr{+0.038}{-0.008}, \quad \lambda=
0.2258\aerr{+0.0005}{-0.0006}, \quad \bar\rho=
0.136\aerr{+0.022}{-0.028}, \quad \bar\eta=
0.402\aerr{+0.015}{-0.054}.
\end{equation}
The constraints are shown in Figure~\ref{Fig:DeltaF2}. The data still
allow sizable NP contributions in both $\Bd$ and $\Bs$ sectors up
to 30--40\% at the 3$\sigma$ level. The results for the CKM
parameters can be compared with those of
Equation~\ref{eq:CKMparams2016}, with the caveat that the inputs
are different. Unsurprisingly, there is a wider range of
variations of the CKM parameters once some of the constraints
involve not only SM but also NP contributions.

The same kind of analysis has also been used for prospective
studies that take into account the accuracies expected from the full
data sets of the LHCb phase 1 upgrade and
Belle-II~\cite{Charles:2013aka}. Assuming no signal of NP, the
constraints on $\Delta_d$ and $\Delta_s$ tighten, setting stringent
constraints on the scale of NP involved, which can range from $10$
to $10^3$ TeV, depending on the structure of couplings chosen.

\boldsubsubsection{Violation of lepton flavour universality in $\Delta F=1$ processes}\label{Sec:globalLFUV}

As discussed in Section~\ref{Sec:LFU}, there are interesting hints
of a breakdown of lepton flavour universality in both $b\to
c\ell\nu$ and $b\to s\ell\ell$ processes. Both types of processes
have been analysed to extract information about potential NP
contributions in the effective Hamiltonian approach describing the
process at the scale $\mu_b=O(m_b)$ around the $b$ quark mass
after integrating out heavier degrees of
freedom~\cite{Buchalla:1995vs}.

For the $b\to c\ell\nu$ transitions, the ratios of the branching
ratios $R(D)$ and $R(D^*)$ do not involve CKM parameters. The
deviations can be easily interpreted by adding new interactions to
the effective Hamiltonian, for instance, additional NP scalar
couplings~\cite{Fajfer:2012vx}. A more extensive
study~\cite{Freytsis:2015qca} highlights a few scenarios
that are compatible not only with the branching ratios but also with the
$q^2$ shape of the $B\to D\tau\bar\nu_\tau$ differential decay
rate. Two-dimensional scenarios with left- and right-handed
couplings, either vector or scalar, are favoured. Note that the $B\to D\ell\bar\nu_\ell$ form factors are known
from lattice QCD simulations~\cite{Na:2015kha,Lattice:2015rga},
but this is not the case for the $B\to D^*\ell\bar\nu_\ell$ decay,
whose prediction requires many additional theoretical assumptions
(validity of heavy-quark effective theory, absence of NP for
electrons and muons). Moreover, presently there are only a very
limited number of observables (two ratios of branching ratios).
The geometry of the decay products could add further information
about the deviations observed in the branching
ratios~\cite{Becirevic:2016hea} and could enable one to check the
$q^2$ dependence of the differential decay rates for both vector
and pseudoscalar final mesons.

There is a much larger set of observables %provided by LHCb and Belle
concerning \decay{b}{s\ellell} decays, with many different
channels. Interest in a global analysis of such decays was
clear long before the advent of \B-factory and LHCb
data~\cite{Ali:1994bf}. The appearance of several tensions in
different $b\to s\ell^+\ell^-$ channels is interesting because all
these observables are sensitive to the same couplings ${\cal
C}_{7,9,10}^{(\prime)}$ induced by the local four-fermion
operators in the effective Hamiltonian approach:

\begin{eqnarray}
\mathcal{O}_9^{(\prime)}=\frac{\alpha}{4\pi}[\bar{s}\gamma^\mu
P_{L(R)}b] [\bar{\mu}\gamma_\mu \mu], &\quad& \mathcal{C}_9^{\rm SM}(\mu_b)=4.07, \\
\mathcal{O}_{10}^{(\prime)}=\frac{\alpha}{4\pi}[\bar{s}\gamma^\mu
P_{L(R)}b] [\bar{\mu}\gamma_\mu\gamma_5\mu], &\quad & \mathcal{C}_{10}^{\rm SM}(\mu_b)=-4.31, \nonumber\\
\mathcal{O}_7^{(\prime)}=\frac{\alpha}{4\pi}m_b[\bar{s}\sigma_{\mu\nu}P_{R(L)}b]F^{\mu\nu},
&\quad & \mathcal{C}_7^{\rm SM}(\mu_b)=-0.29, \nonumber
\end{eqnarray}
where $P_{L,R}$ project on left- and right-handed chiralities and
primed operators have vanishing or negligible Wilson coefficients ${\cal C}^\prime_{7,9,10}$ in the SM. The couplings ${\cal C}_{7,9,10}^{(\prime)}$ can be
constrained through various observables in radiative and (semi-)leptonic $B^0_{(s)}$ decays, each of them sensitive to different
subsets and combinations of coefficients. The first analyses
performed in this spirit and exploiting LHCb
data~\cite{Descotes-Genon:2013wba} pointed to a large contribution
to the Wilson coefficient ${\cal C}_{9}$ in $b\to s\mumu$, which was quickly
confirmed \cite{Altmannshofer:2013foa,Beaujean:2013soa}.
Three recent global
analyses~\cite{Descotes-Genon:2015uva,Altmannshofer:2014rta,Hurth:2016fbr}
have been performed, involving similar sets of data. There are also several analyses that have included the latest observables violating lepton-flavour universality such as $R_{K^*}$~\cite{Ciuchini:2017mik,Geng:2017svp,Hiller:2017bzc,DAmico:2017mtc,Altmannshofer:2017yso,Capdevila:2017bsm}.
They rely on different inputs and hypotheses but agree in their
conclusions and prefer scenarios involving a significant
contribution to ${\cal C}_{9}(m_b)\simeq -1.1$ in $b\to s\mumu$,
whereas contributions to other Wilson coefficients are only loosely
bound and compatible with the SM. Intense theoretical activity
is currently under way to cross-check the various sources of
theoretical uncertainties [power corrections to the limit
$m_b\to\infty$, form factors, long-distance charm-loop
contributions~\cite{Jager:2012uw,Jager:2014rwa,Ciuchini:2015qxb,Descotes-Genon:2014uoa,Ciuchini:2016weo,Bobeth:2017vxj}],
confirming the robustness of this picture up to now.

As there is no clear picture for NP models that could be
responsible for the deviations in both $b\to c\ell\nu$ and
\decay{b}{s\ellell} decays (even though leptoquarks, $Z'$ bosons,
and partial compositeness models are favoured), it is not easy to
perform a combined fit of the CKM parameters and NP contributions in a
way similar to the $\Delta F=2$ case reported in
Section~\ref{Sec:DeltaF2NP}. Indeed, the NP analyses have often
assumed values of the CKM parameters based either on full global
fits or on tree-level determinations, assuming that the uncertainty
coming from the CKM parameters is subleading compared with other
sources of uncertainties.

However, if there is a violation of lepton flavour universality,
all leptonic and semileptonic decays may be significantly
affected. Unfortunately, not all measurements are given for muonic
and electronic modes separately. Removing all these modes from the
determination of the CKM parameters leads to
\begin{eqnarray}
A&=& 0.831\aerr{+0.058}{-0.109}, \quad
\lambda=0.213\aerr{+0.010}{-0.005}, \quad \bar\rho =
0.127\aerr{+0.019}{-0.019}, \quad \bar\eta = 0.350\aerr{+0.012}{-0.011},\\
|\Vcb|&=&0.0421\aerr{+0.0011}{-0.0016}, \quad
|\Vts|=0.0414\aerr{+0.0010}{-0.0016}.\nonumber
\end{eqnarray}
A second approach is also possible, following the current
experimental indications that electron modes are in agreement with
SM. Only the \Pmu and \Ptau modes should be removed from the
global fit to the CKM parameters, leading to
\begin{eqnarray}
A&=& 0.831\aerr{+0.021}{-0.031}, \quad
\lambda=0.2251\aerr{+0.0004}{-0.0004}, \quad \bar\rho =
0.155\aerr{+0.008}{-0.008}, \quad
\bar\eta = 0.340\aerr{+0.010}{-0.010},\\
|\Vcb|&=&0.0425\aerr{+0.0007}{-0.0018}, \quad
|\Vts|=0.0410\aerr{+0.0014}{-0.0012}. \nonumber
\end{eqnarray}
In both cases, $|\Vtb|$ is unity up to a very high accuracy. These
results can be compared with those from the SM global fit
in Equation~\ref{eq:CKMparams2016}:
\begin{equation}
|\Vcb|=0.0418\aerr{+0.0003}{-0.0006}, \quad
|\Vts|=0.0411\aerr{+0.0003}{-0.0006}.
\end{equation}
Removing part or all the modes potentially affected by the
violation of lepton flavour universality significantly
increases the uncertainties (up to a factor of five) on the CKM matrix elements
$|\Vcb|$ and $|\Vts|$, which arise in $b\to c\ell \nu$ and $b\to
s\ell\ell$ decays, respectively. However, considering the other
experimental and theoretical uncertainties involved, the
parametric uncertainty coming from CKM parameters indeed remains 
subleading for the NP analyses of these modes, and it should not
alter their conclusions.

\section{OUTLOOK}\label{Sec:Outlook}

The CKM matrix is a key element in the description of
flavour dynamics in the SM. With only four parameters,
this matrix is able to describe a wide range of phenomena, such
as \CP violation and rare decays. It can thus be constrained by
many different processes, which have to be measured experimentally
with high accuracy and computed with good theoretical control.
After the first LEP measurements, the turn of the millennium has
opened the \B-factory era, leading to a remarkable improvement in
the number and accuracy of the constraints set on the CKM matrix,
which exhibit remarkable consistency and have led to a precise
determination of the CKM parameters.

The status presented in Section~\ref{Sec:Constraints} is based on
experiments up to and including the lifetime of the \B factories,
as well as LHC Run 1. The corresponding data sets have been
almost fully exploited, whereas no updated measurements using data
from the ongoing Run 2 are yet available. This situation will soon
change, as the first Run 2 analyses will shortly be released by LHCb, ATLAS, and
CMS. A change of gear is expected after the year 2020, when both
Belle-II and the phase 1 upgraded LHCb experiment will collect
data at much higher luminosities. The target is a multiplication
of the data sets by up to two orders of
magnitude~\cite{Aushev:2010bq,LHCb-TDR-012}. In the case of LHCb,
this includes the increase of the \bquark{}\bquarkbar
cross section at higher energies and an improved
trigger~\cite{LHCb-TDR-016}. A reduction of experimental
uncertainties by factors of around 10 on the angles $\beta$, $\gamma$,
and $\varphi_s$ is to be expected, as no irreducible systematic
uncertainties are foreseen to affect the results in the
foreseeable future. One may also expect improvements in the experimental measurements of the observables related to the angle $\alpha$ and the matrix elements $|V_{ub}|$ and $V|_{cb}$. In addition, new measurements concerning lepton flavour universality and observables in rare decays are likely to be presented in the coming years.

The interpretation of these improved measurements will depend on developments in
theoretical calculations. The computation using lattice QCD
simulations has already reached a very mature stage for some of
the quantities described in Section~\ref{Sec:global}, for instance,
decay constants and form factors. At the accuracy obtained, some
issues become relevant, such as the estimation of electromagnetic
corrections, the detailed extrapolation in heavy-quark masses, and the
kinematic range available for heavy to light form factors. Hopefully, the resulting improvement in the accuracy of the theoretical computations will resolve the puzzles currently affecting the
determination of $|V_{ub}|$ and $|V_{cb}|$. More generally, the experimental accuracy obtained for the individual constraints requires one to reassess some of the theoretical hypotheses commonly used to extract these quantities and add systematics that have been neglected up to now (e.g., sources of isospin breaking arising in the determination of $\alpha$, penguin pollution for $\beta$). Other improvements can
be expected concerning more exploratory domains, such as the
matrix elements of operators beyond the SM (which are needed to
analyse flavour constraints in NP models) or quantities involving
hadrons difficult to access up to now---for instance, unstable
mesons decaying under the strong interaction (e.g., $\rho$,
$K^*$) or light or heavy baryons (e.g., nucleons, hyperons,
$\Lb$). Progress can also be expected from other
theoretical methods (e.g., effective theories, dispersive
approaches). Even though it is more difficult to assess
their impact on the study of the CKM matrix, these advances should
help in the study of $\epsilon'/\epsilon$, the constraints on NP from neutral-meson mixing, or the interpretation of
anomalies in rare \bquark decays.

The current picture provided by global fits to CKM parameters
within the SM is both accurate and consistent, and it shows that
that this approach can be used to study NP models affecting flavour dynamics (such as models with NP in $\Delta F=2$ transitions). Such analyses extend the initial objective of constraining the CKM matrix, and they require a joint determination of the CKM parameters and NP contributions, based on a larger set of measured observables. Such analyses extend the initial objective of constraining the CKM
matrix, and they require a joint determination of the CKM
parameters and NP contributions. This approach through global fits is currently relevant for the
study of hints of violation of lepton flavour universality in
$b\to c$ and $b\to s$ transitions, which have sparked a great deal of
interest. Several attempts to analyse these deviations in terms
of model-independent effective approaches exist, but these results still need to be connected with viable high-energy models. In these
challenging analyses, the uncertainties related to CKM parameters
are subleading compared with other (experimental
and hadronic) uncertainties. A consistent picture of whether
lepton universality holds may become available soon, which could provide original directions for these studies.

More generally, new developments in flavour physics can be
expected through the improved determination of CKM parameters, the
identification of departures from the SM in flavour
transitions, and the study of heavy degrees of freedom through
low-energy processes at high intensity. In all of these areas,
upcoming measurements from LHCb and Belle-II and ongoing
progress in theoretical computations will play an essential role
in the near future.
\section*{DISCLOSURE STATEMENT}
The authors are not aware of any affiliations, memberships,
funding, or financial holdings that might be perceived as
affecting the objectivity of this review.
\section*{ACKNOWLEDGEMENTS}
S.D.G. thanks his collaborators from the CKMfitter Group for discussions and comments on many issues covered in this review. S.D.G. acknowledges partial support from contract FPA2014-61478-EXP. This work has received funding from the European Union's Horizon 2020 research and innovation program under grant agreements 690575, 674896, and  692194. This work is also part of the NWO Institute Organisation (NWO-I), which is financed by the
Netherlands Organisation for Scientific Research (NWO).

\setboolean{inbibliography}{true}
\bibliographystyle{LHCb-10}
\bibliography{LHCb-PAPER,LHCb-CONF,exp,theory,LHCb,LHCb-TDR}

\end{document}